%% file: vuls-in-ML-systems_v1.tex
\documentclass[sigconf, anonymous = false]{acmart}

\usepackage[nolist]{acronym}
\usepackage[disable]{todonotes}
\AtBeginDocument{%
  \providecommand\BibTeX{{%
    \normalfont B\kern-0.5em{\scshape i\kern-0.25em b}\kern-0.8em\TeX}}}

\copyrightyear{2020} 
\acmYear{2020} 
\setcopyright{rightsretained} 
\acmBooktitle{New Security Paradigms Workshop 2020 (NSPW '20), October 26--29, 2020, Virtual conference}
\acmDOI{10.1145/3442167.3442177}
\acmISBN{978-1-4503-8995-2/20/10}

\acmConference[NSPW '20]{New Security Paradigms Workshop}{October 26--29, 2020}{Virtual conference}

\begin{document}

\title{On managing vulnerabilities in AI/ML systems}

\author{Jonathan M. Spring}
\email{jspring AT sei dot cmu dot edu}
\orcid{0000-0001-9356-219X}
\affiliation{%
  \institution{CERT\textregistered ~Coordination Center\\Software Engineering Institute\\Carnegie Mellon University}
  \city{Pittsburgh}
  \state{PA}
  \postcode{15213}
}
\author{April Galyardt}
\affiliation{%
  \institution{Software Engineering Institute\\Carnegie Mellon University}
  \city{Pittsburgh}
  \state{PA}
  \postcode{15213}
}
\author{Allen D. Householder}
\orcid{0000-0001-8970-4108}
\affiliation{%
  \institution{CERT\textregistered ~Coordination Center\\Software Engineering Institute\\Carnegie Mellon University}
  \city{Pittsburgh}
  \state{PA}
  \postcode{15213}
}
\author{Nathan VanHoudnos}
\affiliation{%
  \institution{Software Engineering Institute\\Carnegie Mellon University}
  \city{Pittsburgh}
  \state{PA}
  \postcode{15213}
}

\renewcommand{\shortauthors}{Spring, Galyardt, Householder, and VanHoudnos}

\begin{abstract}
This paper explores how the current paradigm of vulnerability management might adapt to include machine learning systems through a 
thought experiment: what if flaws in \acf*{ML} were assigned \acp*{CVE-ID}? 
We consider both \acs*{ML} algorithms and model objects.
The hypothetical scenario is structured around exploring the changes to the six areas of vulnerability management: discovery, report intake, analysis, coordination, disclosure, and response. 
While algorithm flaws are well-known in the academic research community, there is no apparent clear line of communication between this research community and the operational communities that deploy and manage  systems that use \acs*{ML}. 
The thought experiments identify some ways in which \acsp*{CVE-ID} may establish some useful lines of communication between these two communities. 
In particular, it would start to introduce the research community to operational security concepts, which appears to be a gap left by existing efforts. 
\end{abstract}



\maketitle

\section{Introduction}
\label{sec:intro}

The topic of this paper is more ``security for automated reasoning'' and less ``automated reasoning for security.''
We will introduce the questions that need to be answered in order to adapt existing vulnerability management practices to support automated reasoning systems. 
We suggest answers to some of the questions, but some are quite thorny questions that may require a new paradigm of either vulnerability management, development of automated reasoning systems, or both. 

First, some definitions. 
We follow the \ac{CERT/CC} definition of vulnerability: ``a set of conditions or behaviors that allows the violation of an explicit or implicit security policy''~\citep[\S1.2]{householder2020cvd}.
We will follow \citet{spring2019ml} and define \ac{ML} as ``a set of statistical tools that analyze data to infer relationships and patterns. Ideally, the relationships and patterns inferred by \ac{ML} will lead to a useful model of the object or phenomenon that the data describes,''
and define \ac{AI} as ``a software agent that takes actions based on its environment.''
To be concrete, this paper will focus on vulnerability management for just \ac{ML}-enabled systems. 

One practical way to think of security services for an \ac{ML} system is via the set of services \iac{CSIRT} might provide, which is produced by \ac{FIRST} and documented by \citet{csirtservices_v2}.
A complete risk management and security perspective on \ac{ML} would include more than the \ac{CSIRT} services framework. 
However, we will work from the assertion that to manage the security of an \ac{ML}-enabled system, all \ac{CSIRT} services will need to be able to handle \ac{ML} systems. 

Specifically, of the \ac{CSIRT} services, we are carving out just vulnerability management for discussion. 
The other services, as well as wider issues such as risk management, all have challenges as well, but we leave them as future work. 

Vulnerability management includes six services~\citep[\S7]{csirtservices_v2}: 
\begin{itemize}
\item    Vulnerability discovery / research
\item    Vulnerability report intake
\item    Vulnerability analysis
\item    Vulnerability coordination
\item    Vulnerability disclosure
\item    Vulnerability response
\end{itemize}
These areas cover a wide range. 
They span the interface between software developers, software users, security teams, and people who find flaws in software. 
There is national infrastructure in multiple countries dedicated to support these services and facilitate communication. 
For example, both the United States and the People's Republic of China have \acp{NVD}.
\ac{FIRST}, a global body, provides one definition of how to score the severity of vulnerabilities, \ac{CVSS}.
The \ac{CVE} scheme is designed to assist such cataloging and ranking efforts.  

\todo{new para}This paper's goal is to facilitate the creation of a trading zone between \ac{ML} engineers, software architects, and security practitioners.
Any trading zone requires a shared language, whether it is a physical or intellectual trading zone \citep{galison1999trading}.
All participants in a trading zone want something of value they can take back to their respective communities.
These items of value are often ``boundary objects,'' which mark boundaries by being recognizable and functional in both cultures in the trading zone. 
\citet{rawls2010thing} states that \ac{MITRE} produces identifiers, such as \acp{CVE-ID}, in part for their value to trading zones as boundary objects. 
This background makes \acp{CVE-ID} an attractive and useful focal point for our thought experiments. 

We will organize our exploration of a new paradigm for \ac{ML} security around one hypothetical -- what if flaws in \ac{ML} systems were assigned \acp{CVE-ID}?
Sections~\ref{sec:VM} and~\ref{sec:VM2} do the main work on exploring the thought experiment.
Before we can answer this question, we first lay some background on the current ways of identifying software vulnerabilities in Section~\ref{sec:Vuls}.
Section~\ref{sec:AML} will provide background on the current state of adversarial attacks on \ac{ML} algorithms. 
Section~\ref{sec:VM} then steps through each of the services areas of vulnerability management to explore the impact of giving ML \emph{algorithm} flaws \acp{CVE-ID}.
Section~\ref{sec:cna_rules} explores the \ac{ML} algorithm thought experiment from the perspective of \acp{cna}. 
Section~\ref{sec:VM2} steps through each of the service areas to explore the impact of giving \ac{ML} \emph{model object} flaws \acp{CVE-ID}.

\section{Vulnerability background}
\label{sec:Vuls}

\todo{edited for clarity}Expanding on the definition of \emph{vulnerability} cited in  Section~\ref{sec:intro}, a vulnerability is ``a set of conditions or behaviors that allows the violation of an explicit or implicit security policy. Vulnerabilities can be caused by software defects, configuration or design decisions, unexpected interactions between systems, or environmental changes''~\citep[\S1.2]{householder2020cvd}.
This definition is useful for the purposes of this paper for clarity, but \ac{CERT/CC} is also a \ac{cna}, so it bears on the ensuing discussion as well. 

\todo{new para}An organization has a variety of options when responding to a vulnerability. 
A fix (a.k.a remediation) is usually defined as a deploying a patch that removes the vulnerable code or retiring the vulnerable system. 
A mitigation reduces the impact of a vulnerability without removing the vulnerable code. 
Example mitigations include adding network segmentation or input and traffic filtering that make it harder to exploit the vulnerability. 
Managing vulnerabilities in \ac{ML} systems will use a combination of remediation and mitigation, just as any other sector.  

There are two common axes that help distinguish vulnerabilities in modern security practice. 
One is within vulnerability identification. 
The second is level of abstraction of the vulnerable product.
Sections \ref{sub:vul-ID} and \ref{sub:vul-abstr}, respectively, discuss these levels.
Section~\ref{sub:CVE} summarizes background the the \ac{CVE} project. 

\subsection{Vulnerability Identification}
\label{sub:vul-ID}

Vulnerability identification and classification spans from scanning
individual systems to organizing vulnerabilities into categories to facilitate better programming principles. A number of vulnerability identification methods exist. \ac{CVE} is perhaps the most widely known, but there are others. 

In increasing broadness along the identification and classification axis, we have: 
\begin{itemize}
\item Instance of a vulnerable product
\item Vulnerability in a product (e.g., \ac{CVE-ID}, \textit{VU\#}) 
\item Category of which a vulnerability is an example (e.g., \ac{CWE}, \ac{OWASP})
\end{itemize}

\subsubsection{Instances of Vulnerable Products}
\label{sub:instances}

On the very specific end of this spectrum, we have instances of a vulnerable product. 
An \emph{instance} is a specific computer or service that uses a vulnerable product.
When an organization scans the systems it owns to perform asset management, it will find instances of a vulnerability. Instances are often tagged as the association of a host or system identifier accompanied by the ID of the vulnerability of which they are an instance.\footnote{Often, this association is mediated through specific versions of software. E.g., host A has version X of software Y installed, and version X of software Y has vulnerability Z} Instances may also be called findings or sightings.

\subsubsection{Vulnerable Products}

Moving up from instances, we find vulnerable products. 
The practitioner community's expectation is that a product is some sort of artificial information processing system.
This definition is vague because a vulnerable product is usually the thing that security practitioners say ``has'' the vulnerability, as defined above.
Since a vulnerability may be introduced by a software defect, configuration decision, design decision, system interaction, or environmental mismatch, the ``product'' that has a vulnerability cannot be constrained much. 
Section~\ref{sub:vul-abstr} will discuss different categories of vulnerable products.

\acp{CVE-ID} are most closely associated with products. 
Section~\ref{sub:CVE} will detail the \ac{CVE} program.

\ac{CERT/CC} publishes Vulnerability Notes using the \textit{VU\#} identifier. 
Like \acp{CVE-ID}, these are also usually at the product level. 
However, while \textit{VU\#} documents often describe a single \ac{CVE-ID}, that is not always the case. 
There are VU\# documents which describe multiple \acp{CVE-ID}, as well as ones that describe vulnerabilities that are out of scope for \ac{CVE} entirely.

Vulnerabilities in products are the main stock and trade of vulnerability management.
Such vulnerabilities often need to be triaged to prioritize actions. 
A popular scoring tool to communicate the technical severity of a vulnerability is \ac{CVSS}.
While \ac{CVSS} and \acp{CVE-ID} are managed by different organizations and are officially unaffiliated with each other, they are often mentally associated due to the close relationship between uniquely identifying and triaging vulnerabilities.
Section~\ref{sub:report-intake} will address \ac{CVSS} in more detail. 

\subsubsection{Vulnerability Categories}

The next broader part is the categorization of vulnerabilities. A number of frameworks exist at this level, with perhaps the best known being the \ac{CWE} framework. 

\ac{CWE} is a "list of common software and hardware weakness types that have security ramifications"~\cite{mitre_cwe}. \ac{CWE} is not intended to be a catalog of specific problems, but rather a collection of important design flaws that lead to ``weaknesses'' in software and hardware.
Although \ac{MITRE} is a bit inconsistent about the definition of ``weakness,'' it is roughly equivalent to the \ac{CERT/CC} definition of vulnerability.
So the two main things that can be members of a \ac{CWE} are a vulnerability and another \ac{CWE} category.  
\acp{CWE} are arranged hierarchically from 10 ``pillar'' weaknesses which are general descriptions of all weaknesses, with intermediate and specific weakness types categorized under them. 

An example \ac{CWE} is ``buffer overflow,'' and any number of \acp{CVE-ID} may be an example of this \ac{CWE}. 
A \ac{CWE} can loosely be understood as a conceptual way that someone might accidentally introduce a security weakness into some information processing system, whereas a \ac{CVE-ID} identifies a concrete product version in which someone introduced a specifically identifiable security flaw.

Not all vulnerabilities associated with a \ac{CWE} get a \ac{CVE-ID}. 
This situation is common with, for example, instances (Section~\ref{sub:instances}) of configuration-level vulnerabilities (see Section~\ref{sub:vul-abstr}) in specific web servers.

The \ac{CWE} specification is ambiguous whether there can be instances of software which match the description of the weakness but cannot have a security impact due to some specific circumstance, such a the code being demonstrably unreachable. 
Various secure coding guidance would certainly recommend avoiding such design patterns because they are fragile \citep{seacord2005secure}. 
This guidance holds whether we name such circumstances a security weakness or not, so pragmatically we shall leave this ambiguity as it is.

\ac{OWASP} is another vulnerability categorization scheme, most famous for its Top 10 document for web developers that ``represents a broad consensus about the most critical security risks to web applications''~\citep{owasptop10}.
Since \ac{OWASP} is tailored to web applications, it is more specific than \ac{CWE}. 
The \ac{OWASP} Top 10 is also more pragmatic; the goal is to prioritize effective protective measures that a web developer should ensure during their development life cycle.  
\ac{OWASP} focuses on secure configuration of web servers, rather than secure coding. 
Only one of the top 10 -- ``9: Using Components with Known Vulnerabilities'' -- overlaps with \acp{CVE-ID}; the other nine represent categories of vulnerabilities that would not normally be given a \ac{CVE-ID}. 

\ac{OWASP} and \ac{CWE} have different constituencies and reach different audiences.
The categorization schemes have differing emphases that reflects their different constituencies. 
But both serve a similar purpose -- to organize knowledge about vulnerabilities in vulnerable products.

\subsection{Abstraction}
\label{sub:vul-abstr}

The second axis, which is independent from vulnerability identification, is a
description of the level of abstraction of the vulnerable product.
The four levels of abstraction for vulnerable products, from most specific to most abstract, are: 

\begin{itemize}
\item Configuration-level vulnerability
\item Implementation-level vulnerability
\item Protocol-level vulnerability
\item Algorithm-level vulnerability 
\end{itemize}

The product may be a specifically-configured instance, an implementation, a protocol, or an algorithm. 

\subsubsection{Configuration vulnerability}

A deployed product may be vulnerable due to its configuration in situ. For example, a linux host may be vulnerable if its `/bin` directory is world-writable due to an errant sysadmin. In such a case there is nothing inherently wrong with the software, it has just been deployed in an insecure manner.

\subsubsection{Implementation vulnerability}
An implementation is, loosely, the source code or binary executable that is distributed as a product. 
Most vulnerabilities that are widely discussed are those found in implemented products, hence \acp{CVE-ID} are most closely associated with implementation vulnerabilities.
The usual way of identifying the vulnerable implementation of a product is to state the versions that are vulnerable, such as ``versions 3.2.9 and earlier are vulnerable.''

\subsubsection{Protocol vulnerability}
Implementations may often be based on a protocol. 
The most common protocol vulnerabilities are in communications protocols -- agreed ways of exchanging information between devices that devices may implement in their own, though mutually compatible, way.
Examples of protocols with documented vulnerabilities include Bluetooth (e.g., CVE-2019-9506), \ac{TLS} (e.g., CVE-2014-3566), and \ac{SMB} (e.g., CVE-2020-0796). 
When there is a protocol vulnerability, all implementations of that protocol are, by definition, vulnerable. 
There may be workarounds to reduce exposure, as usual, but an implementation inherits many things from the protocol it implements, including vulnerabilities.

Vulnerability managers need not localize a vulnerability to a protocol; practically, it is every implementation of the protocol that must change. 
The rules for assigning \acp{CVE-ID} address this directly.
A single ID is assigned to the protocol, standard, or \ac{API} rather than multiple \acp{CVE-ID} assigned to each implementation if and only if ``there is no option to use the functionality or specification [e.g., protocol] in a secure manner''~\citep[\S7.2]{mitre2020cna}.
So in the case where the \ac{TLS} protocol had a vulnerability, every implementation of \ac{TLS} would share the same \ac{CVE-ID}. 
A pragmatic effect of assigning \acp{CVE-ID} to protocols rather than their various implementations is that it makes clear that the protocol designer or standards body is responsible for fixing the vulnerability. 

\subsubsection{Algorithm vulnerability}
The layer of abstraction above protocol is an algorithm vulnerability. 
Historically, this term has usually applied to cryptographic algorithms.
For example, the cryptanalysis of \ac{DES} in the early 1990s \citep{matsui1993des} identified algorithmic vulnerabilities in \ac{DES} that any protocol using that algorithm inherited.
Any implementations of those protocols also inherited the algorithm vulnerabilities as well, as expected. 
Thankfully, vulnerabilities in cryptographic algorithms have become quite rare. 
Such vulnerabilities largely predate the current vulnerability management apparatus of \acp{CVE-ID} which has come to dominate since 2010.
But there is precedent in CVE-2004-2761 for assigning \acp{CVE-ID} for cryptographic weaknesses (in this case, the MD5 algorithm's susceptibility to hash collisions).

Our placement of ``algorithm'' as strictly above ``protocol'' in the abstraction levels is an artifact of the history of networking and network security. 
Communications protocols arrange certain building blocks to reliably and securely exchange information.
A particularly important one of those building blocks is cryptographic algorithms. 
Protocols infrequently but occasionally have vulnerabilities; this is usually a problem in the structure of the protocol and how information is exchanged or handled. 
But the protocol designers usually treated the cryptographic algorithms as special, as a sort of root of trust for the security of the protocol. 
However, the perspective \ac{MITRE} takes with the \ac{CVE-ID} rules would consider both protocols and cryptographic algorithms ``products'' whose functionality would be shared by other products~\citep[\S7.2]{mitre2020cna}. 

\vspace{1em} %

The specificity and abstraction descriptions are orthogonal. 
One can have an instance of a implementation vulnerability, a product with an implementation vulnerability, or an implementation vulnerability which is an example of a weakness type. 
Similarly, one can have an instance of a configuration vulnerability, a protocol (that is, a product) with a specific vulnerability, a vulnerability in an algorithm which is an example of a weakness type, etc. 

This paper will discuss the hypothetical of assigning \acp{CVE-ID} to \ac{ML} algorithm vulnerabilities and/or to \ac{ML} model objects. 
This hypothetical is specifically about vulnerabilities the existing regime does not handle.
The existing vulnerability management regime does not have any problem handling implementation-level vulnerabilities in \ac{ML} libraries, such as buffer overflow mistakes in TensorFlow (e.g., CVE-2018-10055). 
Such algorithm-level vulnerabilities are well known within the \ac{ML} research community, as Section~\ref{sec:AML} will discuss.
However, the current vulnerability management paradigm has not had to handle many algorithm-level vulnerabilities in more traditional computing infrastructure; the last one was probably 2008 with practical collision attacks against the MD5 algorithm~\citep{vu836068MD5}.
This mismatch is one aspect that will make our thought experiments instructive.  

\subsection{CVE-ID background}
\label{sub:CVE}

\acp{CVE-ID} are designed to provide unique identifiers for the purpose of tracking a vulnerability throughout vulnerability management processes, with an emphasis on enabling communication among constituents and stakeholders.
The \ac{CVE} program is not a stand-in for all of vulnerability management: \todo{new 2 sent}there are relevant vulnerabilities that are never assigned \acp{CVE-ID}.
Misconfigured file permissions are a common example.
However, since the \ac{CVE} program provides the unique identifiers that vulnerability managers use to track their main work items, it is a useful entry point that enables our thought experiments to touch, if not fully explore, all six areas of vulnerability management. 

\ac{MITRE} is the lead organization, but they have delegated the ability to assign \acp{CVE-ID} to about 120 \acp{cna}~\citep{mitre_cve}.
The first \acp{CVE-ID} were assigned in 1999, with 1,500 vulnerabilities assigned identifiers that year -- many of which had been discovered some time earlier in the decade. 
As of Aug 19, 2020, about 140,000 vulnerabilities have \ac{CVE} entries. 

\acp{CVE-ID} have power within vulnerability management. 
For example, the US \ac{NVD} requires a \ac{CVE-ID} for all entries (\url{https://nvd.nist.gov/general/FAQ-Sections/CVE-FAQs}).
Because \ac{NIST} operates the \ac{NVD} and \ac{NIST} produces the information security standards for the US federal civilian government, when US government security regulations say something like ``patch all known vulnerabilities,'' the word ``known'' is usually understood to mean ``in the \ac{NVD}.'' 
Which implies that the only vulnerabilities US federal civilian government entities are required to patch are those with \acp{CVE-ID}. 

In the commercial vulnerability management space, a similar scenario plays out. 
Asset management or vulnerability scanning products have a tendency to be based on fingerprints or signatures of device or software stacks.  
For example, if a scanner can determine that a web server is Apache version 2.2.31, then a simple lookup indicates it is vulnerable to CVE-2017-9788 and should be patched to a more recent version. 
As a consequence, vulnerability management is not driven by vulnerabilities so much as it is driven by \acp{CVE-ID}.
The only community in which \acp{CVE-ID} do not entirely drive vulnerability management is website owners, where \ac{OWASP} and \ac{CWE} are used to label configuration-level vulnerabilities such as cross-site scripting and improper data protection configurations.
 
\ac{MITRE} does not strictly control what counts as a vulnerability. 
It is worth quoting their definition at length~\citep[\S7]{mitre2020cna}:
\begin{quote}
The CVE Program does not adhere to a strict definition of a vulnerability. For the most part, CNAs are left to their own discretion to determine whether something is a vulnerability. Root CNAs may provide additional guidance to their child CNAs. This allows the program to adapt to definitions used in different industries, legal regimes, and cultures.

7.1.1 If a product owner considers an issue to be a vulnerability in its product, then the issue MUST be considered a vulnerability, regardless of whether other parties (e.g., other vendors whose products share the affected code) agree.

7.1.2 If the CNA determines that an issue violates the security policy of a product, then the issue SHOULD be considered a vulnerability.

7.1.3 If a CNA receives a report about a new vulnerability that has a negative impact, then the reported vulnerability MAY be considered a vulnerability.
\end{quote} 

Section~\ref{sec:cna_rules} will show this official definition allows space to consider flaws in \ac{ML} algorithms as vulnerabilities that get \acp{CVE-ID}. 
Nothing in the current written guidance would need to change.
However, Section~\ref{sec:VM} will also show how \ac{ML} algorithms present a number of challenges to existing vulnerability management practices, including assumptions about the responsibility to fix \acp{CVE-ID}. 
A trained model object (see Section~\ref{sub:operational-aml}) is fairly clearly a product to which a \ac{CVE-ID} could be assigned; Section~\ref{sec:VM2} will show that choice would present a related but distinct set of challenges to existing vulnerability management practice.

\section{Adversarial ML background}
\label{sec:AML}

There are myriad ways in which an adversary can cause an \ac{ML} algorithm to behave unexpectedly and violate either implicit or explicit security policies. 
Statisticians and \ac{ML} engineers rarely express
such problems in %
vulnerability management terms. 
This section will introduce how the
\ac{ML} %
research, policy, and operational communities have expressed the problem. 

The name of this field is \acf{AML}.
Unfortunately, even here we have a terminology collision; some communities use \ac{AML} to refer to generative adversarial networks, or training \ac{ML} algorithms using game theory through adversarial examples. 
This paper exclusively uses \ac{AML} to refer to attacking and defending \ac{ML} algorithms. 
Section~\ref{sub:academic-aml} summarizes the state of academic \ac{AML} work via the conclusions of two recent literature reviews. 
Section~\ref{sub:policy-aml} summarizes two recent attempts to translate the conclusions out of the \ac{AML} research space to policy makers. 
Section~\ref{sub:operational-aml} introduces modern operational considerations around engineering reliable \ac{ML} systems. 
The understanding of these other efforts maps out empty spaces where a perspective from vulnerability management may be helpful.

\subsection{Summary of academic work}
\label{sub:academic-aml}
\citet{biggio2018wild} is
a highly cited %
literature review within the statistical research community. 
Their abstract summarizes the state of affairs as ``adversarial input perturbations carefully crafted either at training or at test time can easily subvert [\ac{ML} systems'] predictions. The vulnerability of machine learning to such wild patterns (also referred to as adversarial examples), along with the design of suitable countermeasures, have been investigated in the research field of adversarial machine learning.''
While \citet{biggio2018wild} does not use vulnerability management terms, their categorization is based on the basic security triad of confidentiality, integrity, and availability. 
The earliest published work on attacking \ac{ML} algorithms documented by \citep{biggio2018wild} dates to 2004. 

Contemporary with \citet{biggio2018wild}, \citet{papernot2018sok} is an excellent literature review of the space, but whose target audience is the security research community. 
Similar to \citep{biggio2018wild}, \citet{papernot2018sok} find that ``there is growing recognition that ML exposes new vulnerabilities in software systems, yet the technical community's understanding of the nature and extent of these vulnerabilities remains limited.''

Both \citet{biggio2018wild} and \citet{papernot2018sok} taxonimize attacks on \ac{ML} algorithms similarly, though there are certainly differences of emphasis. 
The rest of this section will discuss each of the following aspects of
their %
taxonomies in more detail. 
\begin{itemize}
\item Both distinguish between training-time and test-time attacks. 
\item Both differentiate based on how much the adversary needs to know in order to perform the attack. 
\item They differ slightly on their implied security policies; \citet{papernot2018sok} identifies different kinds of attacks that violate integrity and privacy, rather than the CIA triad \citet{biggio2018wild} uses.  
\item The papers differ in their proposed defenses. 
\end{itemize}

Both distinguish between training-time and test-time attacks. 
If an adversary can influence the data used to train a model, different attacks are possible than if the adversary can influence the data items to be tested. 
The situation where an attacker can influence both test and training is often called \emph{poisoning}, whereas the situation is called \emph{evasion} if just test data can be influenced~\citep[\S3.3]{biggio2018wild}. 
\citet{biggio2018wild} restrict their framework to supervised learning algorithms.
\citet[\S5]{papernot2018sok} notes the training-test distinction is heavily biased towards just supervised classification algorithms, but indicates that other types of algorithms, such as unsupervised and reinforcement learning, seem to exhibit similar vulnerability even though they are much less thoroughly studied.  
Beyond algorithm type restrictions, this distinction covers only the model building and validation life cycle and excludes model deployment~\citep{galyardt2020comments}.
Restricting the scope to model building and validation makes sense for academic research, but our \ac{CVE-ID} thought experiment will need to include deployment and environmental vulnerabilities, as discussed in Section \ref{sub:operational-aml}.

Both differentiate based on how much the adversary needs to know in order to perform the attack. 
The terms ``white box'' (full-access) and ``black box'' (query-access) are used with similar meaning in both papers, and their meanings are similar to the way the terms are used in the fuzzing literature~\citep{manes2019art}.\footnote{\label{footnote-terms}These terms based on color were commonly used in the literature, but we will use the more descriptive and less divisive terms ``full-access'' and ``query-access'' in their place in our discussion.}
Briefly, full-access refers to the attacker having complete access to the model object, such that the attacker can load it into a controlled environment and inspect and modify its components.
Query-access commonly refers to the attacker being able to provide inputs to the model and received outputs, but not be able to inspect the internals of the model object.  

In broad strokes, if the adversary knows more about the model and the feature space, it is easier for them to attack the model. 
However, query-access attacks on models are readily feasible.
More precisely, algorithms ``can be threatened without any substantial knowledge of the feature space, the learning algorithm and the training data, if the attacker can query the system in a black-box manner and get feedback'' \citep[\S3.2]{biggio2018wild}. 
At the extreme end, in special cases an attacker can create a full-access situation from a query-access situation by recreating the model (up to machine precision!) through querying the model with carefully crafted examples \cite{carlini_cryptanalytic_2020}.  

Although \citet{papernot2018sok} and \citet{biggio2018wild} differ slightly on their implied security policies, both use the term ``threat model'' to discuss the adversary's capabilities and goals. 
This is part of a security policy, but falls far short of defining a security policy for an \ac{ML} system. 
Within the \ac{AML} literature, ``threat model'' is used similar to the mathematical cryptography community, where a threat model is a mathematical expression of the adversary's capabilities. 
In \ac{AML}, a threat model is a declarative statement. 
In operational security communities, a threat model is a the output of an investigation or observations about what adversaries can in fact do, or have done in the past, see for example \citet{fox2018threat}. 
The difference between a declarative and observational threat model is subtle, but it may cause members of the two communities to miscommunicate.

One consequence of this disconnect is that the \ac{AML} community has adopted threat models that are mathematically tractable, but not likely to be observed in practice. 
For example, the most popular \ac{AML} threat models are based on small changes to the input, usually measured with an $L_p$ norm. 
These are useful for two reasons: first, they tractable to analyze, and second, they allow researchers to develop a principled understanding of the fundamental properties of these systems. 
These threat models, however, are divorced from the constraints of the real world. 
Specifically, permissions are often implemented on computer systems to be all or nothing; if an adversary has write access to a file, they can make arbitrary changes. 
If the threat model assumes the adversary has some form of write access, as implied by the ability to make changes to the input, then bounding the scope of the changes the adversary would choose to make is somewhat implausible.

Portions of the \ac{AML} community consider more realistic threat
models. 
An important class of these are modifications to the physical environment that must survive a data processing pipeline, such as a sticker that would cause an object to be misclassfied, poisoning the data that one would collect for training, or inserting trojans or backdoors into publicly released models.
However, these communities are a minority within \ac{AML} and have not yet received the same amount of attention as those communities that rely on mathematically tractable threat models that were the focus of the literature reviews. 
Whether such mathematically bounded threat models apply to these physical environments is unclear; if they do, it is likely only in the context of a wider system where such limits are either empirically motivated or enforced by other security mechanisms (such as human guards).

The literature review papers also differ in their proposed defenses.
\citet{biggio2018wild} recommends reactive defenses, security by design, and security by obscurity. 
``Reactive defenses'' are similar to what a security practitioner might call ``continuous monitoring and evaluation'' in conjunction with an appropriate risk assessment \citep[p.11]{spring2019ml}.  
\citet{papernot2018sok} presents a more conservative analysis of defenses, emphasizing that defense against most known attacks is an open area of research. 
Therefore, \citep{papernot2018sok} prioritizes mapping out a research plan for a science of security and privacy of \ac{ML}.\footnote{``Science of security'' is a broad term, and it is not clear which of the senses or goals identified by \citet{spring2017why} is meant by \citep{papernot2018sok}.} 
The broad recommendation is that \ac{ML} algorithms will need to become resilient to distribution drifts and incorporate privacy through differential privacy methods. 
While this perspective essentially admits researchers do not currently know how to defend \ac{ML} algorithms, that perspective should sit better with the security community than uncritically endorsing security through obscurity.

\subsection{Summary of policy work}
\label{sub:policy-aml}

This section is not a comprehensive survey of policy work related to \ac{AML}, but rather highlights two ongoing projects that are attempting to bridge the gap between \ac{ML} researchers and software engineers or policy makers. 
The first is a private sector initiative, led by Microsoft and Harvard.
The second is an effort by \ac{NIST} to bridge the \ac{AML} and information security policy communities. 

The Microsoft/Harvard effort is perhaps centered on \citet{kumar2019failure}, but includes other collaborative documents such as \citet{kumar2018law}.
\todo{new ref / new 2 sent}\citet{kumar2020adversarial} specifically names tracking, scoring, and responding to vulnerabilities in \ac{ML} systems as a gap in current practice when \iac{ML} system is under attack. 
This paper fleshes out what it would take to fill that gap.

One important goal stated by \citet{kumar2019failure} is to ``equip software developers, security incident responders, lawyers, and policy makers with a common vernacular to talk about'' \ac{AML} through a taxonomy of \ac{ML} failure modes.
The taxonomy expands beyond the narrow research concerns of the academic literature, and so includes deployment and environmental failures. 
However, the \citet{kumar2019failure} taxonomy remains restricted by current research in some ways, in that attacks on supervised classification algorithms are better researched and better understood than failures of other types of algorithms; the authors attempt to overcome this gap and be as comprehensive as plausible. 

\citet{kumar2019failure} adopt some taxonomic categories also used in the academic literature reviews. 
The distinction between full-access and query-access attacks (see footnote~\ref{footnote-terms}) is present here with similar meaning. 
The connection to security policy is via the CIA triad, similar to \citet{biggio2018wild}. 
\citet{kumar2019failure} frames CIA as assurances for the \ac{ML} system, rather than capabilities of the adversary, which aligns better with common security usage.\footnote{There are shortcomings with this common usage. One one hand, it is based on a metaphor with the physical goods, such as fortifications, which breaks down when applied to information systems~\citep{pieters2011social}. On the other hand, some standards bodies, such as the \ac{IETF}, list several other recommended security services in additional to just CIA, such as non-repudiation, and differentiate some parts of CIA, such as between system integrity and data integrity~\citep[p.274]{rfc4949}. Failure modes in \ac{ML} may put additional pressure on the basic CIA triad, and justify a change to one of these more nuanced approaches.} 

\citet{kumar2019failure} identifies 11 intentionally motivated failure modes, namely:
\begin{enumerate}
\item Perturbation attack 
\item Poisoning attack
\item Model inversion
\item Membership inference
\item Model Stealing
\item Reprogramming ML system
\item Adversarial example in physical domain
\item Malicious ML provider recovering training data
\item Attacking the ML supply chain
\item Backdoor ML
\item Exploit Software Dependencies
\end{enumerate}
In relation to the \ac{CVE} paradigm of vulnerability management, the best way to understand these failure modes is as new candidate \acp{CWE}. 
The failure modes are about general kinds of things that can go wrong when designing, implementing, or using \ac{ML} algorithms. 
Also similar to \ac{CWE}, \citet{kumar2019failure} is clear the authors are curating a living document that will change as the community finds further failure modes. 

\todo{new para}\ac{MITRE} has published one CWE related to \ac{AML}: CWE-1039, ``Automated Recognition Mechanism with Inadequate Detection or Handling of Adversarial Input Perturbations.''
CWE-1039 tracks to ``Perturbation attack'' in \citet{kumar2019failure}. 
This suggests there are at least 10 further CWEs to define. 
CWE-1039 is more general than either of our proposed thought experiments -- either a specific algorithm or a specific model could be an example of this weakness type.  

\ac{NIST} has a policy effort to bridge \ac{AML} and information security policy in \citet{nistir8269}. 
The NISTIR is a draft, and so will likely change and improve; at present, a comment period closed on Jan 30, 2020 and the authors are editing based on those comments. 
The version presently available is essentially just an attempt to synthesize the academic surveys; these include \citep{biggio2018wild, papernot2018sok} as well as three others as primary sources \citep[p.2]{nistir8269}. 

Despite being a \ac{NIST} document, one thing that \citet{nistir8269} decidedly does not do is integrate or deconflict \ac{AML} terminology with the terminology in the \ac{NIST} Computer Security Research Center glossary, or any other set of standard information security terms~\citep{galyardt2020comments}.
Unlike \citet{kumar2019failure}, the NISTIR draft is narrowly scoped to academic \ac{AML} interests and does not account for deployment or environmental failures that would be of interest in practice \citep{galyardt2020comments}. 
We expect the next draft will improve on these shortcomings. 
However, for these reasons at present \citet{nistir8269} does not bridge the \ac{NVD}, \todo{edited for clarity}\ac{CVE} program, or other vulnerability management functions with the \ac{AML} space.

\subsection{Summary of operational work}
\label{sub:operational-aml}

The work on operational assurance that \ac{ML} systems behave as expected covers a broad space. 
First, this section introduces an anatomy of an operational \ac{ML} system to clearly define the parts subject to our thought experiments and contextualize them. 
Second, we will scope our thought experiment by summarizing valid operational concerns that are and are not security vulnerabilities through the lens of deployment context.

\begin{figure*}
    \centering
    \includegraphics[trim=10 30 0 0,clip,width=0.9\textwidth]{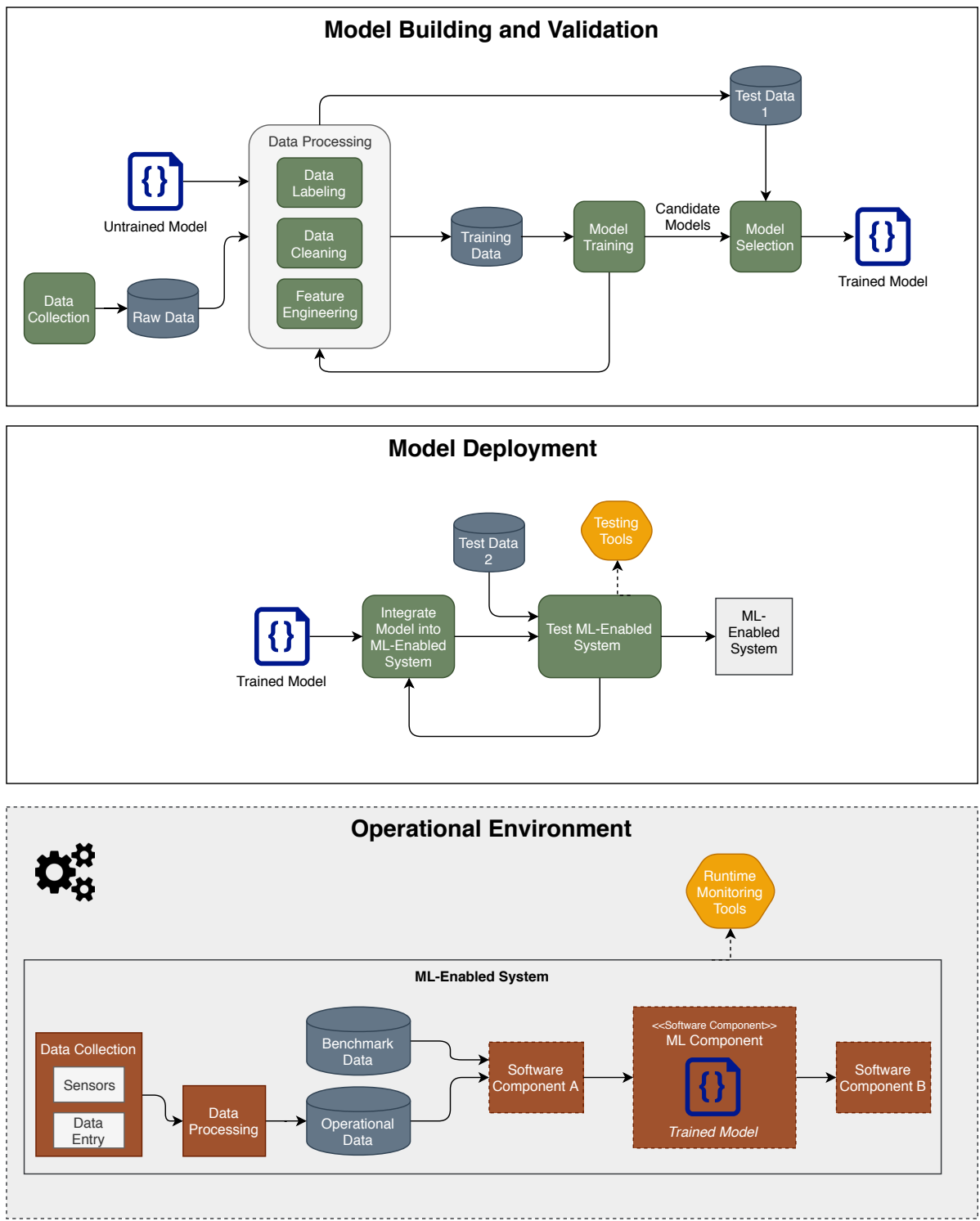}
    \caption{Each of the green boxes represents a process, often with a human directly involved. Each of the burnt-orange boxes represents a software component in the operational environment. In an operational ML system, the model needs frequent updating, and this process could be represented by adding a loop to this diagram.}
    \label{fig:MLPipeline}
\end{figure*}

A deployed ML system has a broader attack surface than just the training and testing of the ML model. %
Our thought experiment in this paper is limited in scope to \ac{ML} algorithms and model objects, but it is important to recognize that these are just some parts of an operational \ac{ML} system. 

Figure \ref{fig:MLPipeline} provides a representation of the pieces involved in developing and operating an ML system.\footnote{Note that this diagram can represent both supervised and unsupervised ML systems; however for reinforcement learning, the general pipeline is the same, but the model building and validation stage will require modification.} While this diagram is only a rough representation of any particular system, it provides a useful tool for conceptualizing the possible vulnerabilities of any deployed ML system. Each of the components and processes in the diagram represents a different point where a vulnerability could be introduced; 
vulnerabilities may arise in sensors that collect the data, the data processing component, or the runtime monitoring tools, in addition to the model itself.

This perspective highlights the difference between an \ac{ML} algorithm and an \ac{ML} system. The \ac{ML} algorithm is the particular mathematical procedure used to create an \ac{ML} model object by configuring/training it on the data. The \ac{ML} system includes all of the components illustrated in Figure \ref{fig:MLPipeline}. 
\citet{horneman_ai_2019} provides general guidance for the design and management of \ac{ML} systems. 

As discussed in Section \ref{sub:academic-aml}, the \ac{AML} research community focuses on categories of vulnerabilities introduced during 
the Model Building and Validation stage; particularly, categories of vulnerabilities that may be inherent in the Untrained Model 
or introduced by manipulation of the Training Data or Test Data 1. 
\citet{kumar2019failure}, discussed in Section \ref{sub:policy-aml}, broadens this focus; e.g., the category "Attacking the ML supply chain" introduces the idea that a vulnerability could be introduced to a trained model as it is being downloaded in the Model Deployment stage. 
However, a deployed system has yet more components which could introduce vulnerabilities. 
For example, if an adversary wanted you to waste time and money retraining your model, thus hurting you in a resource-constrained environment, they could attack the Benchmark Data to make your model seem as if performance was degrading for unknown reasons. 

To situate \ac{ML} systems in their full context, one should observe that deployed \ac{ML} systems often are part of enforcing or developing policies for organizations, including governments. 
Such systems often embed power structures, biases, and inequity \citep{zou2018ai,oneil2016weapons,noble2018algorithms}.
Both \ac{ML} researchers, e.g., \citep{Farnadi18, Russell15}, and legal scholars, e.g., \citep{Citron14, Eidelman18, Kroll17, Lehr17}, have been struggling with how to seek out and eliminate these problems from \ac{ML} systems. However, designing an ML system with an assured fidelity to a particular substantive policy choice is not usually considered a security question. That is, assuring a high-quality \ac{ML} system generally involves assuring attributes other than security, such as fairness, equity, stability, etc.
We place these other attributes out of the scope of this paper. 

However, exceptions for these other attributes aside, the larger context of deployment is essential for defining what constitutes a vulnerability. 
Vulnerabilities live at the intersection of the software {\em system} and its interaction with the environment. 
For example,  \citet{Householder2015steampunk} has argued that prior to the invention of airplanes, buildings did not have airplane related vulnerabilities. Something about the world changed with the invention of airplanes, and now the interaction between buildings and their world contains a new class of vulnerable states, damage caused by impacts with airplanes. 
In the same way, before the advent of self-driving cars, \ac{ML} systems did not have vulnerabilities involving stickers on stop signs~\citep{gu_badnets_2019}. 
Since the deployment context has changed to include self-driving cars, the interaction with the environment now includes such vulnerabilities.
Accounting for these diverse environments is both a challenge and a motivation for our thought experiments. 

\section{Thought experiment 1: Algorithms}
\label{sec:VM}

This section explores one hypothetical situation -- what if flaws in \ac{ML} algorithms were assigned \acp{CVE-ID}? 
We explore the consequences for vulnerability management by analyzing each of the six vulnerability management service areas \citep{csirtservices_v2}. 
\ac{CERT/CC} has published one vulnerability note about an \ac{ML}
algorithm vulnerability, but it is tracked only with the internal
identifier VU\#425163 and does not list a \ac{CVE-ID}~\citep{vu425163}.
The note describes how \ac{ML} classifiers trained via a gradient descent algorithm are vulnerable to arbitrary misclassification attacks. 
Although the focus of our thought experiment will tend to be generic,
in cases where a concrete example is helpful we will use the
\ac{CERT/CC} vul note. 

As Section~\ref{sub:operational-aml} discussed, there are many parts that go into an operational \ac{ML} system, and the algorithms are just one part. 
The thought experiment here is just working through what would happen if \acp{CVE-ID} were assigned to algorithms.
Of course, systems that use the algorithm should likely inherit the \ac{CVE-ID}; although there is guidance for defending \ac{ML} systems from vulnerabilities in their algorithms, the guidance is evolving rapidly because most guidance has been quickly shown to be inadequate \citep{carlini2019evaluating}.

Each of the following subsections examines our thought experiment through one of the \ac{CSIRT} services within vulnerability management, as introduced in Section~\ref{sec:intro}.

\subsection{Vulnerability discovery / research}

In our scenario, vulnerability discovery will not change much.
All current vulnerability discovery work can continue as before.
If we start assigning \acp{CVE-ID} to \ac{ML} algorithm
vulnerabilities the number of people who should be counted as doing
vulnerability discovery will jump higher overnight, as \ac{AML} folks
will be added to the ranks.
Although it is difficult to estimate the size of the \ac{AML}
community, over 1,900 papers have been published since 2014 according
to a widely followed tracker \citep{carlini_nicholas_complete_2020},
suggesting a substantial influx.

As in the current paradigm, in our thought experiment not all \todo{Clarity edit} ML algorithm vulnerabilities will end up with an assigned \ac{CVE-ID}.
However, the reasons for this will remain the same; there are at least three possible reasons why a \ac{CVE-ID} may not be assigned:
\begin{itemize}
\item The vulnerability research is internal to product development, and will be fixed without being made public
\item The vulnerability is not independently fixable from another vulnerability
\item The vulnerability researcher plans to sell the vulnerability or otherwise use it in attacks
\end{itemize}
 The first two are expressly forbidden to have \acp{CVE-ID} in the CVE specification \citep[\S7]{mitre2020cna}.
Those in the third category may be assigned a \ac{CVE-ID} eventually, once defenders detect attacks using it. 

This first reason will not unduly stress our thought experiment. 
Most mature software engineering projects conduct internal testing before release. 
Tests may include fuzzing, program verification, code review, or unit tests.
In general, if code pre-release fails these tests, it is fixed without being assigned \iac{CVE-ID} because it does not need to be discussed publicly.
Such testing regimes for \ac{ML} systems are one recommendation in both the academic and policy literature.

The second reason a vulnerability may not receive \iac{CVE-ID} is more slippery in relation to \ac{ML} algorithms. 
Given that how to prevent these attacks on \ac{ML} algorithms is not known in principle, it may be considered an area of active research.
However, we can say something about what should get different \acp{CVE-ID}.
If the algorithms are distinct \ac{ML} algorithms, then they should get distinct \acp{CVE-ID}, even if they exhibit the same flaw type (as documented by \citep{kumar2019failure}, for example). 
This practice would be analogous to the way in which two web browsers get separate \acp{CVE-ID} even if they both have a buffer overflow, even though buffer overflow is the same kind of flaw in both. 
Similarly, if there are distinct versions of the same algorithm, they will only get separate \acp{CVE-ID} if the fixes are independent.
This practice is analogous to the way that multiple versions of a web browser may be affected by a single \ac{CVE-ID} today. 

The abstraction level hierarchy presented in Section~\ref{sub:vul-abstr} presents some further trouble for ``independently fixable'' that is not so easily resolved. 
The hierarchy implies that all protocols or models that use a vulnerable algorithm and implementations of those protocols or models will share the same \ac{CVE-ID}. 
This sharing is implied because the implementations are not independently fixable from each other; the algorithm may need to be fixed and then this change propagated down the abstraction levels. 
However, as is the case currently where every instance of a product is not always susceptible to vulnerabilities in the product, due to configuration changes or other workarounds in place, it may be possible to use a vulnerable algorithm in a way that that does not expose every algorithm vulnerability. 
This, too, is an area of active research, and a topic we will return to in Sections~\ref{sub:v-analysis} and~\ref{sub:response}.  

Finally, vulnerabilities discovered for the purpose of attacking others would mostly have the same properties in a world where \ac{ML} algorithms get \acp{CVE-ID}.
It may seem that disambiguating different attacks is harder with \ac{ML} algorithms. 
But given the current prevailing exploit-as-a-service~\citep{Grier:2012:MCE:2382196.2382283} and crimeware-as-a-service~\citep{sood2013crimeware} situations, it is already exceedingly difficult to know which vulnerability an adversary exploited.   
And if \ac{ML} algorithms get \acp{CVE-ID}, it may slowly become easier to identify and catalog which attacks an \ac{ML} system is expected to be vulnerable to or resist.

\subsection{Vulnerability report intake}
\label{sub:report-intake}

In addition to receiving reports, this service area is where any analysts ``review, categorize, prioritize, and process'' a report~\citep{csirtservices_v2}. 
There are various human processes that will need to adjust if \ac{ML} algorithms receive \acp{CVE-ID} and become part of this process.
We will focus on prioritization. 
Review, categorization, and processing will all require analysts to acquire new skills and expertise to understand \ac{ML} algorithms, but this may be handled in the medium term by creating a \ac{CSIRT} dedicated to the open-source \ac{ML} community. 
Such specialized \acp{CSIRT} exist for industrial control systems, for example. 
And for broader security awareness beyond vulnerability management, \acp{ISAC} serve a similar function. 
Prioritization cannot be so easily handed off to a specialized workforce. 

Prioritization implies giving \iac{CVSS} score to the vulnerability~\citep{cvss_v3-1}. 
Although a \ac{CVSS} base score is explicitly not to be used on its own as a prioritization score, the technical severity as measured by \ac{CVSS} is often considered an  important factor in vulnerability prioritization for many organizations today.  \todo{Clarity edit here} The \ac{CVSS} base score is intended to reflect the ``reasonable worst case impact'', and encapsulates the relative ease of exploitation (exploitability) and the direct consequences of a successful exploit (impact). 
The dimensions of exploitability include attack vector, attack complexity, required privileges, and user interaction; while impact considers scope and the confidentiality-integrity-availability triad \citep[\S1]{cvss_v3-1}. 
Table~\ref{tab:cvss-base}\todo{New table} summarizes the options for \ac{CVSS} base metrics. 

\begin{table}
    \centering
    \begin{tabular}{@{} l|l| p{.33\columnwidth} @{}}
        Metric Name & Group & Options \tabularnewline \hline
        Attack Vector & Exploitability & Physical; Local; Adjacent; Network \tabularnewline
        Attack Complexity & Exploitability & Low; High \tabularnewline 
        Privileges Required & Exploitability & None; Low; High \tabularnewline 
        User Interaction & Exploitability & None; Required \tabularnewline
        Scope & Impact & Changed; Unchanged\tabularnewline 
        Confidentiality  & Impact & None; Low; High \tabularnewline 
        Integrity  & Impact & None; Low; High \tabularnewline
        Availability  & Impact & None; Low; High\tabularnewline 
    \end{tabular}
    \caption{Summary of CVSS v3.1 base metrics \citep{cvss_v3-1}.}
    \label{tab:cvss-base}
\end{table}

What would happen if we try to give a \ac{CVSS} score to  VU\#425163? The first thing we notice is that the guidance of evaluating the ``reasonable worst case impact" has limited utility when applied to ML algorithms. 
The meaning of ``reasonable'' in the context of an algorithm is difficult to bound, as an algorithm could be used in so many different contexts.
When considering exploitability, there are plausible scenarios where the misclassification can be forced remotely (attack vector: network), the attack can be automated~\citep{papernot2016technical} (attack complexity: low), the adversary can submit test cases without prior authentication (privileges required: none), and the user does not have to do anything (user interaction: none). 
\todo{Clarity edit here}For impact, the first metric is scope -- can the compromise of the vulnerable component be used by the adversary to compromise other parts of the information system of which it is a part to increase the scope of the attack. 
For ML algorithms, attacking the behavior of the ML algorithm can change the behavior of the car, robot, or system in which the trained ML model is embedded (scope: changed). 
The last three questions are about the CIA triad, but since scope is changed it is the worse case of the algorithm itself or the wider scope of what is the adversary can access. 
\citet{kumar2019failure} would characterize this kind of misclassification attack as a failure of algorithm integrity, so we can safely set integrity to high.
The same \ac{CVE-ID} will apply to forcing a Tesla autopilot to change lanes, which, in a reasonable worst case, could cause the car to crash, which sets availability to high as well. 
Even if we ignore confidentiality and set it to none, the \ac{CVSS} score is maxed out at 10.0. 

This is probably not a reasonable \ac{CVSS} score, or at least it is out of sync with scores for other vulnerabilities. 
It is unclear whether this indicates our \ac{ML} algorithm vulnerability is particularly bad, or whether \ac{CVSS} just is not suited to analyze such vulnerabilities. 
Given that \ac{CVSS} has a host of problems, including complaints it does not apply properly to domains such as industrial control systems, medical devices, and robots, we hypothesize the problem lies with \ac{CVSS} in our thought experiment as well. 
\citet{spring2020ssvc} proposes an alternative, stakeholder-specific vulnerability prioritization scheme that would be more easily adapted to \ac{ML} algorithm vulnerabilities.

\subsection{Vulnerability analysis}
\label{sub:v-analysis}

The purpose of vulnerability analysis is to triage incoming reports, understand the root cause of the vulnerability, and develop countermeasures to fix or mitigate the vulnerability~\citep{csirtservices_v2}.

Triage is heavily dependent on prioritization, and so mirrors the discussion in Section~\ref{sub:report-intake}. 
The ingredient triage adds is asset management. 
For example, if the vulnerable component identified by the \ac{CVE-ID} is not installed anywhere the security team is responsible for, then the priority does not matter.
Asset management of \ac{ML} algorithms is a challenging problem. 
The modern cybersecurity paradigm already struggles with vulnerabilities in libraries that may be used in diverse products; these challenges are well-documented by the work on \ac{SBOM}~\citep{manion2019sbom}.
Tracking vulnerable \ac{ML} algorithms will only make this problem more urgent. 
Such tracking may meet initial resistance from the system vendors, under the argument that \acp{CVE-ID} may reveal the algorithm they use and that choice may be intellectual property or some such. 
Whether this legal argument carries will likely be jurisdiction specific; in places with a strong right to explanation (such as in the GDPR), such information seems less likely to be protected. 

Root cause analysis is affected in similar ways as vulnerability discovery. 
The \ac{AML} academic research field is currently engaged in root cause analysis for all the vulnerable algorithms. 
What assigning \acp{CVE-ID} would change is to drive conversation between the \ac{AML} research community and system engineers and system owners, who will likely gain increased awareness that their \ac{ML} systems have flaws. 
As \citet{kumar2019failure} identifies, there is a gap here in that the two communities lack a shared language in which to communicate.
So far the main efforts seem to have been to teach system owners the language from \ac{AML}. 
Assigning \acp{CVE-ID} to algorithm vulnerabilities would start the work of teaching the \ac{AML} researchers language from operational security. 

Developing countermeasures for vulnerabilities in \ac{ML} algorithms will continue to be a hard problem. 
While there is 15 years of research within the \ac{AML} community, we are not aware of any public discussion on how to build an \ac{ML} system with appropriate mitigations or workarounds in place to isolate the algorithm from adversary interference.
Security professionals use mitigations and workarounds for vulnerabilities regularly, when a fix is not available or not practical. 
For example, one recommended mitigation for any \ac{SMB} service is to only expose it to a trusted local network, never the internet. 
This workaround does not fix \ac{SMB} vulnerabilities, but it does mitigate them.  
\citet[\S5.1]{biggio2018wild} summarizes six papers under the heading ``reactive defenses'' that likely map to workarounds and mitigations. 
\acp{CVE-ID} may help system owners and developers track which workarounds are necessary for their systems. 
But the system engineering practices for \acs{ML}-enabled systems are still a work in progress as well~\citep{horneman_ai_2019}. 
The need is especially dire if such systems have a cybersecurity task such as malware identification, which must be exposed to adversary-crafted input~\citep{spring2019ml}.

\subsection{Vulnerability coordination}
\label{sub:coordination}

\ac{CVD} is ``the process of gathering information from vulnerability finders, coordinating the sharing of that information between relevant stakeholders, and disclosing the existence of software vulnerabilities and their mitigations to various stakeholders, including the public''~\citep[\S1.2]{householder2020cvd}.
The previous sections have highlighted the knowledge gap between security operations and \ac{AML} researchers. 
\ac{CVD} is probably the place where those knowledge gaps will evidence as barriers. 
All three steps -- gathering information, sharing that information, and disclosing vulnerabilities -- are dependent on shared knowledge. 

For example, consider VU\#425163. 
If that vulnerability were to get a \ac{CVE-ID}, its finders in the \ac{AML} community would likely react with confusion.
The issue has been known for years; they would have some questions about why assign an identifier now. 
Next, who do we share information about the vulnerability with -- it is already public, after all. 
And there is no easy way, at present, to notify specific vendors because there is no listing of which products use which algorithms. 
Despite these facts, the statements from the engineering and policy community are that the information about these flaws is not getting where it needs to; that is the whole premise of the efforts by \cite{kumar2019failure}.
Communication and coordination seem to be failing somewhere, likely at multiple points. 
Assigning VU\#425163 a \ac{CVE-ID} would not solve any of these problems.
But it might start some conversations that may build some trust and shared understanding that form the beginnings of coordination. 

\subsection{Vulnerability disclosure}
\label{sub:disclosure}

Disclosure is closely linked to coordination. 
Two topics that would be affected here are announcements and timelines.
Vulnerability announcements are affected because they include the results of several things
discussed above, such as severity scores, identifiers, recommended fixes or mitigations, and a description using shared terminology.   
The relevant constituents may also be different from current vulnerability announcements, as Section~\ref{sub:coordination} indicated.
While these are a lot of changes to announcements, none of them require further discussion; timelines do.

In security operations, vulnerability disclosure timelines are already a contentious topic. 
The consensus position is often described as ``coordinated,'' as in \ac{CVD}, where a vendor is told about vulnerabilities in their software before the public or attackers so that the vendor has a chance to develop and deploy a fix. 
The \ac{AML} community works under what might be called a zero notification paradigm -- results tend to be published with no prior warning to those who develop or use the algorithm. 
There are security practitioners who advocate and practice this for vulnerabilities in traditional systems, but it is not the norm in most communities. 
These two differing sets of norms would come into conflict if a widely used and distributed product were tagged with a \ac{CVE-ID} because overnight an \ac{ML} algorithm vulnerability is discovered, posted to arXiv, and assigned a \ac{CVE-ID}.
How this conflict would be resolved depends on various political, operational, and technical factors. 
While we cannot venture a prediction as to how the conflict will resolve, the fact that there will be a conflict between these two norms is almost certain. 

\subsection{Vulnerability response}
\label{sub:response}

Vulnerability response is where operations folks do something to prevent vulnerabilities from being exploited.
There are two basic steps: detecting which systems an organization manages are vulnerable to which flaws and applying fixes or mitigations to those systems.
\todo{new 3 sent}For any system, traditional IT or \ac{ML}, some of the vulnerable systems an organization manages will not have \acp{CVE-ID}; web server misconfigurations are a common example. 
We focus on those vulnerabilities with \acp{CVE-ID}.
Both detection and response would need to adapt if \ac{ML} algorithms are assigned \acp{CVE-ID}, because in practice many detection and response workflows are based on \acp{CVE-ID} as the primary unit of work.  

The change to detection would depend on how much work can be done during analysis and coordination to link a vulnerable algorithm to deployment in specific product versions. 
This connection gets easier if the vulnerability is associated with certain published models that use the algorithm, but may get harder if the vulnerability depends on features of the training or test data.
However, proprietary products rarely reveal their model or algorithm, which, in the short term at least, will make detection challenging. 

It is likely that there will not be simple vulnerability scans to detect \ac{ML} algorithm vulnerabilities, which upends the current detection paradigm that relies heavily on operations like Nessus scans. 
If detection does not adapt, one possible outcome of our thought experiment is that while \ac{ML} algorithms have \acp{CVE-ID}, there is no operational impact because those \acp{CVE-ID} are never identified on deployed systems. 
For at least the medium term, detection of systems with vulnerable \ac{ML} algorithms would be manual or mostly manual based on annotation of asset management databases. 
Whether or not this is acceptable will depend on the volume of vulnerable systems and the volume of attacks against systems without fixes or mitigations in place. 

As Section~\ref{sub:v-analysis} discussed, developing fixes and mitigations for these algorithm vulnerabilities will continue to be a hard problem. 
Response has a dependency on there being a fix or mitigation for operations folks to apply.
This alludes to another possible fate of the thought experiment: there is raised awareness of exactly which systems are vulnerable to which kinds of attacks, but nothing for anyone to do about it. 
This statement is a bit overly dramatic, of course.
System owners can either make a risk management decision that the system provides enough value to be worth the risk, or not. 
And usual system security principles such as least access and least privilege should still apply. 

A more measured possible fate of our thought experiment is that \emph{system owners become aware that they have taken on more vulnerable systems than they expected or understood, and re-evaluate either their need or deployed protections for those systems}. 
Until the \ac{AML} community can provide more comprehensive fixes, this may be the best response available. 
And, if we establish the connection to between \ac{AML} research and operational security sooner, then it should reduce the time to communicate and deploy those fixes when they become available.  

\section{By the letter of the rules}
\label{sec:cna_rules}

Section~\ref{sec:VM} explored the impact of assigning \acp{CVE-ID} to \ac{ML} algorithm vulnerabilities in general. 
This section explores the details of the \ac{cna} rules on assigning \acp{CVE-ID} to specifically ask whether \ac{CERT/CC} could be justified in assigning \iac{CVE-ID} to the vulnerability note VU\#425163 identifying gradient descent as vulnerable to misclassification attacks \citep{vu425163}. 
The \ac{cna} rules \citep{mitre2020cna} have four aspects to consider:
\begin{enumerate}
    \item What is a Vulnerability?
    \item How many Vulnerabilities?
    \item \ac{cna} Scope
    \item Requirements for Assigning a CVE ID 
\end{enumerate}

We consider each of these in detail below.

\subsection{What is a Vulnerability?}

\ac{CVE-ID} assignment rules~\citep{mitre2020cna} allow for a degree of latitude for \ac{cna} judgement, but do provide some specific guidance. 
Each of these criteria can be applied to
VU\#425163, though it takes a bit more work than with a traditional implementation-level vulnerability in a product. 

Rule 7.1.1 says if the vendor recognizes the report as a vulnerability, then the report must be considered a vulnerability. 
In the case of most protocols and some algorithms, there is often a standards body responsible for maintaining the specification. 
That standards body is seen as the vendor for the purposes of \ac{CVE-ID} assignment.  
For example, the cryptographic algorithm MD5 is specified in \ac{IETF} RFC-1321~\citep{rfc1321} and has at least one assigned \ac{CVE-ID} (CVE-2004-2761). 
But unlike MD5, the gradient descent algorithm is not "owned" by a standards body.
There are multiple variants of gradient descent~\citep{ruder2016overview}, but the basic stochastic algorithm for training neural networks dates to the late 1980s~\citep{becker1989improving}.
Therefore it is difficult to pin down a specific "vendor" who would authoritatively judge 7.1.1.
The authors of the paper are the closest thing to the role of "vendor", but this seems like a poor fit for assigning responsibility for maintenance. 
Vulnerabilities can still be identified in abandonware though, so this should not be a big problem.

Rule 7.1.2 says the report should be considered a vulnerability if a product security policy is violated. 
In the case of a stochastic algorithm such as a classification algorithm, it seems that acceptable false positive or false negative rates might constitute such a policy. 
If an attacker can present the system with inputs that would otherwise be rare, it seems that a policy based on FPR or FNR would be violated. 
One might make a similar argument about fit quality for a regression needing to meet some specified tolerance.
But stochastic policies can be difficult for security analysts to reason about. 

The question of negative impact posed in Rule 7.1.3 is somewhat harder.
The assertion of VU\#425163 is that all systems in which gradient descent is used are susceptible to exploitation by adversarial input.
So even though some implementations might not have a security relevant impact were exploitation attempted, it seems likely that \textit{some} security relevant impact will occur in \textit{some} implementations.
So far, it seems our strongest arguments in favor of treating VU\#425163 as a vulnerability per \ac{CVE-ID} assignment rules are 7.1.2 and 7.1.3 coupled with \ac{cna} discretion.

\subsection{How many Vulnerabilities?}

Proceeding through the four questions, we reach section 7.2 of the \ac{cna} rules, which focuses on the concept of independent fixes. 
Rule 7.2.1 simply prohibits duplicate assignments. Rule 7.2.2 prohibits assignment when a dependency on another vulnerability exists, which is not the case for our example. 
Rule 7.2.3 suggests resolving uncertainty about independence by assigning a single \ac{CVE-ID}. In effect, splitting is taken to be easier than merging should revisions be needed.

Rule 7.2.4 addresses what to do when ``multiple products are affected by the same independently fixable vulnerability'' arising from shared code. 
But both protocols and algorithms can have multiple implementations and therefore may not share code even though they share a vulnerability, making this rule inapplicable.

Rule 7.2.5 resolves the situation when the vulnerability originates in functionality or specification, as is the case for both protocol and algorithm vulnerabilities.
If there is a way to implement the functionality securely, then each implementation that fails to do so must get its own \ac{CVE-ID} according to rule 7.2.5.a.
Conversely, a single \ac{CVE-ID} is required by rule 7.2.5.b when there is no way to implement the specification or functionality securely.
Rule 7.2.5.c resolves ambiguity in favor of multiple assignments.

\todo{new para}In understanding 7.2.5, compare the vulnerability in this gradient descent algorithm to known errors in floating point handling algorithms. 
There is not a vulnerability in floating point handling per se because it is possible to handle floating points properly.
Floating point algorithms have known errors that can lead to vulnerabilities, for example, CVE-2006-6499. 
This situation is an example of 7.2.5.a. 
The C Secure Coding standard discusses these floating point algorithm issues under FLP00-C through FLP07-C \citep{seacord2005secure}.
However, there is no known secure method for training a model with gradient descent (see Section~\ref{sub:academic-aml}). 
So it does not seem comparable to algorithms like floating point handling -- which have known problems but also have known secure methods for use. 

VU\#425163 describes a problem with every system that uses gradient descent in training models, so rule 7.2.5.b seems most relevant here, thereby requiring a single \ac{CVE-ID} assignment rather than one per affected product. 
This is consistent with assigning \acp{CVE-ID} to cryptographic algorithms and protocol specifications as previously noted. 

\subsection{Scope and Requirements}
\label{sub:CNA-scope}

The remainder of the \ac{CVE-ID} assignment rules are easier to get through: Rule 7.3 verifies that the vulnerability is in scope for the \ac{cna} making the assignment. 
\ac{CERT/CC}'s scope as a \ac{cna} covers assignment related to its vulnerability coordination role, so this falls within our scope. 

Rules 7.4.1, and 7.4.2 verify that the report is intended to be public, which is true because VU\#425163 exists.
Rule 7.4.3 prohibits duplicate assignment to previously assigned \acp{CVE-ID}.
Rules 7.4.4, 7.4.5, and 7.4.6 address the differences between vulnerabilities for which someone other than the \ac{cna} must take action to resolve. 
The only relevant one for our case is rule 7.4.6, which allows assignment for cases where the affected product(s) or service(s) are not owned by the \ac{cna} but are customer controlled. 
\ac{CERT/CC} does not own the gradient descent algorithm, and it is used in customer controlled systems.

Rule 7.4.7 requires assignments not be made for products that are neither licensable nor publicly available. 
For VU\#425163, although the gradient descent algorithm itself is not licensed, it is publicly available, so rule 7.4.7 does not impede us.

Finally, rule 7.4.8 requires \acp{cna} to consider \textit{only} these rules when making assignment decisions. 
Therefore per the \ac{cna} assignment rules, it seems that VU\#425163 deserves a CVE ID.

\subsection{Assignment conclusions}

In a discussion among the authors and the analyst staff at \ac{CERT/CC}, several of us hold the view that VU\#425163 and issues like it might be better suited as a category of vulnerability -- such as \ac{CWE} entries or an \ac{OWASP} item to avoid -- rather than \acp{CVE-ID}.
The lack of vendor ownership of an algorithm (or family of algorithms) was one recurring concern.
\todo{new}
This question of ownership is open around ML systems generally; for example, what exactly can be patented (and therefore owned) is unclear. 
But in general, a specific model for a specific purpose can be patented but an algorithm like logistic regression cannot.
If an object is patented, it definitely has an owner; but an unpatentable item may still have an ``owner'' in the CVE sense if, for example, an algorithm is specified by an open standards body. 
We have not been able to resolve the relevant algorithm ownership question to our satisfaction. 

Another concern centered on the fact that not all trained models are exposed to attacker-controlled input to the same degree, so the fact that gradient descent was used to train a model embedded in the system may not imply that an attacker can exploit it.
Finally, from a vulnerability management operational perspective, many organizations have policies that require known vulnerabilities (that is, those with \acp{CVE-ID} assigned) to be fixed in a timely manner.
Because the only known fix for VU\#425163 basically boils down to defense-in-depth, it was easier for many analysts to consider this as a weakness -- and therefore deserving of a \ac{CWE} entry -- rather than a \ac{CVE-ID}.

This situation does not resolve a final recommendation for or against assigning a \ac{CVE-ID} to VU\#425163.
On the one hand, the \ac{cna} rules recommend assigning a \ac{CVE-ID}.
On the other hand, at least some professional vulnerability analysts think of it as a weakness and not a vulnerability. 
These tensions should be addressed in either outcome; it is unclear whether the \ac{cna} rules should change or the professional community intuitions should adapt.

\section{Thought experiment 2: Model~objects}
\label{sec:VM2}
This section explores a second hypothetical situation, if algorithm vulnerabilities such as VU\#425163 are not assigned \acp{CVE-ID}, what if the the trained model objects themselves were assigned \acp{CVE-ID}?

Specifically, as discussed in Section~\ref{sub:operational-aml}, the trained model object within the ML system can enter a vulnerable state when the ML system interacts with its environment. Since the trained model object is a component of the software system that has a defined version, can persist for long periods of time, and can enter a vulnerable state, we explore the consequences of assigning trained model objects \acp{CVE-ID}. 
\todo{para new from here}It is possible this scenario has happened; CVE-2019-8760 identifies a vulnerability in Apple's Face ID software that was fixed by ``improving Face ID machine learning models.''
However, our thought experiment will cover \ac{ML} models generally, not just models with a security function. 
Furthermore, Apple has not released details about its response to CVE-2019-8760, therefore to achieve a useful level of detail in the discussion we will exercise another thought experiment about known flaws in a popular model object.  

As the specific example, consider \citet{xu_adversarial_2019} which generated an adversarial t-shirt pattern that successfully evaded the person detection capability of two COCO \cite{lin_microsoft_2015} trained object detectors, Faster R-CNN \cite{ren_faster_2016} and YOLO v2 \cite{redmon2016yolo9000}. Although \citet{xu_adversarial_2019} do not release their code nor specify precisely which trained model objects they used, there is sufficient detail in their paper that, at a minimum, the torchvision implementation of Faster R-CNN, available in a version pre-trained on COCO \cite{torchvision_fasterrcnn_resnet50_fpn_coco}, and the darknet implementation of YOLOv2, available in a version pre-trained on COCO \cite{yolov2_weights}, are both vulnerable to the adversarial t-shirt pattern. 
For brevity, we refer to this candidate \ac{CVE-ID} as CVE-tee. 

\todo{new para}The hypothetical CVE-tee will focus on these pre-trained model objects. 
However, this example makes space for further questions about to what \iac{CVE-ID} should be assigned. 
YOLOv2 is a framework for training a neural network on image data sets. 
In the terms of Figure~\ref{fig:MLPipeline}, it is a model building and validation framework.
Some aspects of YOLOv2 are fixed -- the \ac{ML} algorithm it trains is a single neural network. 
Many are configurable. 
Some aspects, such as the training data set, are configurable. 
The CVE-tee example focuses on the COCO data set.
There are examples of problems localizable to the training data set -- for example, \citet{buolamwini2018gender}.
Whether \iac{CVE-ID} might be more productively assigned to the model building framework or the training data set are open questions for future work. 
This paper is focused on just two of the more extreme points in the possible space of where \iac{CVE-ID} might be assigned -- the \ac{ML} algorithm (Section~\ref{sec:VM}) and a specific trained model object (this section). 

Each of the following subsections examines our thought experiment through one of the \ac{CSIRT} services withing vulnerability management, following our thought experiment in Section~\ref{sec:VM}.

\subsection{Vulnerability discovery / research}

As in the prior thought experiment, vulnerability discovery will not change much beyond the substantial increase in the number of persons who should be counted as doing vulnerability discovery. For example, there are a growing collection of academic papers that already take as a starting place a publicly available model object and publish exploits that force those model objects into undesirable states, such as our motivating example for CVE-tee. 

As in the current paradigm, not all vulnerable model objects will be assigned a \ac{CVE-ID}, for reasons much the same as in the prior thought experiment. A particular concern for assigning trained model objects \acp{CVE-ID} is persistence of the the trained model. For example, some versioned models might be so short-lived as to not warrant \ac{CVE-ID} assignment, such as during training itself or during internal development. The \ac{CVE} rules do not specifically address a minimum time of existence to assign an ID, although there is mention of the need for public notification, which implies mass human-scale response times. That said, there are many trained model objects that have a persistence measured in years, notably the models released by popular deep learning frameworks, such as in torchvision and keras.io. The YOLOv2 exemplar models for CVE-tee, for example, has been available since 2016. 

A second concern is that, as of this writing, all trained models are vulnerable to an attacker who is aware of the defense strategies used to defend the model \cite{carlini2019evaluating, tramer_adaptive_2020}. This implies that every machine learning system that is released would have open \acp{CVE-ID} associated with its trained model. Such mass assignment is not necessarily a negative thing, because it makes clear to both the machine learning community and the security community that using these models in situations that impact a given security policy is delicate matter that requires careful thought. 

\subsection{Vulnerability report intake}
\label{sub:report-intake2}

In the prior thought experiment, the focus was on assigning algorithms \acp{CVE-ID}.
Section~\ref{sec:VM} has an underlying assumption that there are relatively few algorithms to study. 
In this thought experiment, we consider assigning \acp{CVE-ID} to trained model objects; this suggests several orders of magnitude more \acp{CVE-ID} be assigned because one widely used algorithm can produce many such vulnerable model objects. 

For example, we identified two candidate models for CVE-tee, the official torchvision and darknet implementations of Faster R-CNN and YOLOv2, respectively. 
We chose those models out of familiarity. 
However, there are many additional model files available for download that have been pre-trained on COCO. For example, Faster R-CNN was published at Neurips in 2015, and their initial implementation was in MATLAB (\url{https://github.com/ShaoqingRen/faster_rcnn}). 
The authors later released a version in python (\url{https://github.com/rbgirshick/py-faster-rcnn}), and the success of their approach has led to multiple implementations in pytorch, tensorflow, keras, etc.
These model object versions may be considered similar enough in some important sense to receive the same \ac{CVE-ID}.
However, tracking all of the Faster R-CNN and YOLOv2 model objects trained on COCO and their derivatives is a non-trivial task.

An additional concern, as in the prior thought experiment, is prioritization of an ML vulnerability. 
Although the trained model is a more concrete product than an algorithm, there is still a wide variety of contexts in which the trained models might be used. 
The ``worst case'' reasoning of \ac{CVSS} will lead to us giving this vulnerability, and presumably most \acp{CVE-ID} in trained models, a ``critical'' \ac{CVSS} score (above 9.0). 
While this may be justified, it will cause friction during report intake if there is a large influx of high priority vulnerabilities.

\subsection{Vulnerability analysis}
\label{sub:v-analysis2}
As discussed in the prior thought experiment, root cause analysis and the development of mitigations are open research questions in the AML community. 

The processes of understanding the root cause and developing countermeasures may be aided by having a more concrete model object and vulnerable state to focus on.
For example, in the case of CVE-tee, the vulnerability report is a demonstration that two popular object detectors are fooled by a person wearing a particular pattern on their clothing, specifically a t-shirt. 
This leads to various mitigation strategies that may be more or less appropriate given the broader context of why an object detector is used to detect the presence of persons in a frame of video. 
These could range from social interventions, such as requiring persons to wear a specific uniform and enforcing that requirement without machine learning, to technical ones. 
A particular technical approach might be to move away from standard cameras and object detectors trained on COCO to infra-red cameras and object detectors trained on IR data, such as \url{https://www.flir.com/oem/adas/adas-dataset-form/}. 
Since the engineering involved in creating a thermally active adversarial pattern is harder than printing a t-shirt, this may be a useful mitigation. 

\subsection{Vulnerability coordination}
\label{sub:coordination2}

As in the prior thought experiment, the \ac{AML} community will likely react with confusion if \iac{CVE-ID} is assigned to a set of trained model objects, for much the same reasons as the confusion that would result from an algorithm being assigned \iac{CVE-ID}. 
Again, the general issue has been known for years, the vulnerabilities are already public, and the notification of vendors is difficult because there is no list of which vendors use which model objects (or their derivatives) for their products.
Assigning CVE-tee a \ac{CVE-ID} would not solve any of theses problems, but it might start some conversations that may build the trust and shared understanding necessary for coordination to begin. 

Assigning \acp{CVE-ID} may also have a chilling effect on the willingness of researchers to share trained models with the public. 
If, for example, the assignment of a \ac{CVE-ID} is perceived by the researcher as a negative mark upon their work, then this may make a given researcher less likely to release the code and trained models that make it easier for others to continue to move the field forward or put the results of research to innovative uses. 
Similarly, since the current state of \ac{AML} is unable to isolate the root causes of a given vulnerability -- that is, we do not know what lines of code to change or different algorithm to use -- researchers may perceive the assignment of \acp{CVE-ID} as unnecessary, which may erode rather than build trust. 

Risks of chilling effects notwithstanding, there is reason to believe that the \ac{AML} and security communities can build the necessary relationships. 
First, the \ac{CSIRT} community has been here before. 
For example, the medical device community is on a path to more mature vulnerability management. 
Roughly, this has meant building relationships between the \ac{CSIRT} and medical device communities.  
That journey has included \ac{FDA} regulations on pre-market and post-market handling of vulnerabilities and the Health \ac{ISAC} creating a community in which vulnerability management skills can be cultivated and encouraged \citep{hisacCVD}.
Such relationship building has had fits and starts, but it provides historical lessons that could facilitate improved outcomes for connecting the \ac{AML} and \ac{CSIRT} communities.
Second, there are high-profile examples within the AML community that attempt to perform \ac{CVD}, such as OpenAI staging the release of its GPT-2 model over a period of six months (\url{https://openai.com/blog/gpt-2-6-month-follow-up/}).
This community building will be hard, and regulation is not the right solution for every community, but \ac{ML} is not going to become less important and so it is probably wise to start this hard work. 

\subsection{Vulnerability disclosure}
\label{sub:disclosure2}

As in the prior thought experiment, there is likely to be a conflict between the zero notification paradigm commonly practiced by the \ac{AML} community and the \ac{CVD} paradigm adopted by other security communities. 
This is not ameliorated by changing the level at which the \ac{CVE} is assigned. 
It remains to be seen how this conflict will resolve.
Specifically, it remains unclear if academic researchers, who are under significant pressure to share their results as quickly and widely as a possible, would be willing to wait to publish their exploits. 

\subsection{Vulnerability response}
\label{sub:response2}

The response portion is where operations takes some action to prevent vulnerabilities from being exploited. 
As in the prior thought experiment, the steps are similar: identify which systems are vulnerable to which flaws, and then respond by mitigating those vulnerabilities as necessary. 
The modifications necessary to these processes when ML algorithms are assigned \acp{CVE-ID} are broadly similar to when model objects are assigned \acp{CVE-ID}. 
In both cases, it will be challenging to identify which systems have which vuls, either because the system was trained with a vulnerable algorithm or contains a model object that has a known vulnerability. 

We believe the key difference will be in developing mitigations. 
For example, a \ac{CVE-ID} assigned to VU\#425163 could give only general advice. For example, Figure~\ref{fig:MLPipeline} indicates the adversarial pattern could be dealt with at a variety of levels. 
An operator could  modify the environment so that the sensor is unlikely to encounter the adversarial pattern, or modify the sensor itself to make the pattern more difficult to produce, or add a run-time monitoring tool focusing on detecting such patterns, or modify the software components upstream from the trained model to filter out such patterns, or modify the software components downstream of the trained model to ameliorate the effects of fooled inputs. 
In contrast, if \iac{CVE-ID} were assigned to a model object -- and a particular threat to it, such as CVE-tee -- the advice given could be more specific. 
For example, suggested mitigations could include modifying the social environment to enforce clothing norms that preclude such a pattern, investing in infrared sensors so that an attack would need to produce thermal patterns -- which is much harder -- to fool the sensor, etc. 
Such increased focus and reduced scope may lead to more productive conversations between the relevant stakeholders.

\section{Conclusion}
\label{sec:conclusion}

These changes to vulnerability management are simultaneously minor and revolutionary. 
Although there are important practical differences between assigning \ac{CVE-ID} to algorithms versus model objects, the two thought experiments result in similar changes to the current  vulnerability management paradigm. 

From the following perspectives, the changes are minor.
\begin{itemize}

\item \ac{MITRE}'s guidance for \acp{cna} need not change.
\item The number of \ac{ML} algorithm vulnerabilities would only be a small percentage increase over the 20,000 \acp{CVE-ID} assigned annually (in 2019); model object vulnerability assignments would create more \acp{CVE-ID} than algorithms but still not more than several hundred per year. 
\item The presence of these vulnerabilities in existing \ac{ML} algorithms and model objects is well-known and repeatedly demonstrated.
\item Within vulnerability management and security management more generally, it is normal for sector-specific groups (such as \acp{ISAC} or \acp{ISAO}) to form to handle sector-specific issues. 
\item Challenges in vulnerability management due to supply chain and asset management may be emphasized, but are not new. 
\item Current \ac{CVSS} scoring norms struggle to adapt to various existing stakeholder communities, and it is not clear that \ac{AML} is worse than other areas such as medical devices.  

\end{itemize}

On the other hand, the following changes indicate a paradigm shift for either \ac{AML}, vulnerability management, or both. 

\begin{itemize}

\item Although cryptographic algorithms have had vulnerabilities in the past, algorithm-level vulnerabilities have been essentially unknown in the current vulnerability management regime. 
\item How to fix the vulnerable algorithms or defend model objects is not known, so any assigned \acp{CVE-ID} would be open issues for an unknown length of time.
\item Asset owners and security policy folks who are fluent in vulnerability management do not currently speak a shared language with academic \ac{AML} researchers. 
\item Engineering guidance for \ac{ML} systems is nascent~\citep{horneman_ai_2019} and it is not clear how to handle supply chain documentation or asset management, including whether algorithms should be identified as assets; model objects are often not identified as assets in practice, though it would be easier to identify them than algorithms.
\item Current \ac{CVSS} scoring norms probably are not suitable for either algorithm or model-object vulnerabilities. 
\item \ac{AML} publication timelines and \ac{CVD} norms are in conflict. 

\end{itemize}

Many of these changes are in tension; sometimes one aspect of a change is minor while from another perspective the same change is revolutionary. 
The thought experiment does not provide a strong recommendation for or against; like many problems in vulnerability management, it is nuanced and complicated. 
However, many of the major changes are probably things that would benefit both communities. 
So while they may be hard, they are desirable.
From this perspective, the \ac{ML} engineering, vulnerability management, and \ac{AML} communities probably should build the appropriate bridges and communication infrastructure. 
Then the question is whether we can make the time. 

\begin{acks}
\input{acks.tex}
\end{acks}

\bibliographystyle{ACM-Reference-Format}
\bibliography{work2019,work-20150722,jono,ucl-20190326,nspw,april,adh}

\input{acronyms.tex}

\end{document}

%% file: acks.tex
Copyright 2020 ACM.
This material is based upon work funded and supported by the Department of Defense under Contract No. FA8702-15-D-0002 with Carnegie Mellon University for the operation of the Software Engineering Institute, a federally funded research and development center.
References herein to any specific commercial product, process, or service by trade name, trade mark, manufacturer, or otherwise, does not necessarily constitute or imply its endorsement, recommendation, or favoring by Carnegie Mellon University or its Software Engineering Institute.
[DISTRIBUTION STATEMENT A] This material has been approved for public release and unlimited distribution.  Please see Copyright notice for non-US Government use and distribution.
CERT Coordination Center\textregistered is registered in the U.S. Patent and Trademark Office by Carnegie Mellon University.

%% file: acronyms.tex
\begin{acronym}[CERT/CC]
        \acro{ACL}{Access Control List}
        \acro{ACM}{Association for Computing Machinery}
    	\acro{ACoD}{\acroextra{Art into Science: }A Conference for Defense}
    	\acro{AES}{Advanced Encryption Standard}
	\acro{AI}{artificial intelligence}
    	\acro{AirCERT}{Automated Incident Reporting}
	\acro{AML}{adversarial machine learning}
    	\acro{API}{Application Programming Interface}
    	\acro{APWG}{Anti-Phishing Working Group}
    	\acro{ARMOR}{Assistant for Randomized Monitoring Over Routes}
    	\acro{ARPA}{Advanced Research Projects Agency\acroextra{, from 1972--1993 and since 1996 called \acs{DARPA}}}
    	\acro{attack}[ATT\&CK]{Adversarial Tactics, Techniques, and Common Knowledge\acroextra{ (by \acs{MITRE})}}
    	\acro{BCP}{Best Current Practice\acroextra{, a series of documents published by \acs{IETF}}}
    	\acro{BGP}{Border Gateway Protocol}
    	\acro{BI}{logic of bunched implications}
    	\acro{BIS}{Department for Business, Innovation, and Skills\acroextra{ (United Kingdom)}}
    	\acro{BLP}{Bell-Lapadula\acroextra{, a model of access control}}
    	\acro{C2}{Command and Control}
    	\acro{CAE}{Center of Academic Excellence}
    	\acro{CAIDA}{Center for Applied Internet Data Analysis\acroextra{, based at University of California San Diego}}
    	\acro{CAPEC}{Common Attack Pattern Enumeration and Classification\acroextra{ (by \acs{MITRE})}}
    	\acro{CCIPS}{Computer Crime and Intellectual Property Section\acroextra{ of the \ac{DoJ}}}
    	\acro{CCE}{Common Configuration Enumeration\acroextra{ (by \acs{NIST})}}
    	\acro{CCSS}{Common Configuration Scoring System\acroextra{ (by \acs{NIST})}}
    	\acro{CEE}{Common Event Expression\acroextra{ (by \acs{MITRE})}}
    	\acro{CERT/CC}{CERT{\small{}{\textregistered}} Coordination Center\acroextra{ operated by Carnegie Mellon University}}
    	\acro{CIA}{Central Intelligence Agency\acroextra{ (\acs*{US})}}
    	\acro{CIS}{Center for Internet Security}
	\acro{cna}[CNA]{\ac*{CVE} Numbering Authority}
	\acroplural{cna}[CNAs]{\ac*{CVE} Numbering Authorities}
        \acro{CNA}{Computer Network Attack}
    	\acro{CND}{Computer Network Defense}
    	\acro{CNO}{Computer Network Operations}
    	\acro{CPE}{Common Platform Enumeration\acroextra{ (by \acs{NIST})}}
    	\acro{CSIR}{Computer Security Incident Response}
    	\acro{CSIRT}{Computer Security Incident Response Team}
    	\acro{CTL}{Concurrent Time Logic}
	\acro{CVD}{Coordinated Vulnerability Disclosure}
    	\acro{CVE}{Common Vulnerabilities and Exposures\acroextra{ (by \acs{MITRE})}}
    	\acro{CVE-ID}{\ac*{CVE} identifier}
    	\acro{CVRF}{Common Vulnerability Reporting Framework}
    	\acro{CVSS}{Common Vulnerability Scoring System\acroextra{, maintained by \acs{FIRST}}}
    	\acro{CWE}{Common Weakness Enumeration\acroextra{ (by \acs{MITRE})}}
    	\acro{CWSS}{Common Weakness Scoring System\acroextra{, maintained by \acs{MITRE}}}
    	\acro{CybOX}{Cyber Observable Expression\acroextra{, maintained by \acs{MITRE}}}
    	\acro{DARPA}{Defense Advanced Research Projects Agency}
	\acro{DES}{Data Encryption Standard}
    	\acro{DHS}{\acs*{US} Department of Homeland Security}
    	\acro{DNS}{Domain Name System}
    	\acro{DoD}{\acs*{US} Department of Defense}
    	\acro{DoJ}{\acs*{US} Department of Justice}
    	\acro{ENISA}{\acs*{EU} Agency for Network and Information Security}
    	\acro{EPSRC}{Engineering and Physical Sciences Research Council\acroextra{ (United Kingdom)}}
    	\acro{EU}{European Union}
    	\acro{FAA}{Federal Aviation Administration\acroextra{ (\ac{US})}}
	\acro{FBI}{\acroextra{\acs*{US} }Federal Bureau of Investigation}
	\acro{FDA}{\acroextra{\acs*{US} }Food and Drug Administration}
    	\acro{FIRST}{Forum of Incident Response and Security Teams}
    	\acro{FISMA}{Federal Information Security Management Act\acroextra{ (\acs*{US})}}
    	\acro{FS-ISAC}{Financial Services \acf{ISAC}}
    	\acro{FTP}{File Transfer Protocol}
    	\acro{GCHQ}{Government Communications Headquarters\acroextra{ (United Kingdom)}}
    	\acro{GFIRST}{Government \acs{FIRST}}
    	\acro{HotSoS}{Symposium on the Science of Security}
    	\acro{HTTP}{Hypertext Transfer Protocol\acroextra{, a standard by \acs{W3C}}}
    	\acro{HTCIA}{High Technology Crime Investigation Association}
    	\acro{IC}{intelligence community}
    	\acro{ICT}{information and communications technology}
    	\acro{IEEE}{Institute of Electrical and Electronic Engineers}
    	\acro{IEP}{Information Exchange Policy}
    	\acro{IETF}{Internet Engineering Task Force}
    	\acro{IDS}{Intrusion Detection System}
    	\acro{IODEF}{Incident Object Description Exchange Format}
    	\acro{iodefplus}[IODEF+]{Incident Object Description Exchange Format Extensions\acroextra{ (RFC~5901)}}
    	\acro{IDMEF}{Intrusion Detection Message Exchange Format\acroextra{ (RFC~4765)}}
    	\acro{ISAC}{Information Sharing and Analysis Center}
    	\acro{ISAO}{Information Sharing and Analysis Organization}
    	\acro{ISC}{Internet Storm Center\acroextra{part of the privately-run \acs{SANS}}}
    	\acro{ISO}{International Organization for Standardization}
	    \acro{ISP}{Internet Service Provider}
    	\acro{ITU}{International Telecommunications Union\acroextra{, an agency of the \acs{UN}}}
    	\acro{LAX}{Los Angeles International Airport}
    	\acro{LBNL}{Lawrence Berkeley National Laboratory}
    	\acro{MAEC}{Malware Attribute Enumeration and Characterization\acroextra{ (by \acs{MITRE})}}
    	\acro{MITRE}{the Mitre Corporation}
	\acro{ML}{machine learning}
    	\acro{MMDEF}{Malware Metadata Exchange Format}
    	\acro{MoD}{Ministry of Defence\acroextra{ (United Kingdom)}}
    	\acro{NATO}{North Atlantic Treaty Organization}
	    \acro{NCA}{National Crime Agency\acroextra{ (UK)}}
    	\acro{NCCIC}{\acroextra{\acs*{US} }National Cybersecurity and Communications Integration Center}
    	\acro{NDA}{non-disclosure agreement}
    	\acro{NIDPS}{Network Intrusion Detection and Prevention System}
    	\acro{NIST}{National Institute of Standards and Technology\acroextra{, part of the \acs*{US} Department of Commerce}}
	    \acro{NSA}{National Security Agency\acroextra{ (\acs*{US})}}
    	\acro{NSF}{National Science Foundation\acroextra{ (\acs*{US})}}
	\acro{NVD}{National Vulnerability Database}
    	\acro{OCIL}{Open Checklist Interactive Language\acroextra{ (by \acs{NIST})}}
    	\acro{OVAL}{Open Vulnerability and Assessment Language\acroextra{ (by \acs{MITRE})}}
    	\acro{OWASP}{Open Web Application Security Project}
    	\acro{OWL}{Ontology Web Language}
    	\acro{pDNS}{passive \acf{DNS}\acroextra{ traffic analysis}}
    	\acro{RAM}{Random Access Memory}
    	\acro{RCT}{Randomized Controlled Trial}
    	\acro{REN-ISAC}{Research and Education Networking \acf{ISAC}}
    	\acro{RID}{Real-time Inter-network Defense}
    	\acro{RISCS}{Research Institute in Science of Cyber Security\acroextra{ (United Kingdom)}}
    	\acro{RFC}{Request for Comments\acroextra{, standardization and informational documents published by the \ac{IETF}}}
    	\acro{SANS}[SANS Institute]{Sysadmin, Audit, Network, and Security Institute}
	\acro{SBOM}{Software Bill of Materials}
    	\acro{SCAP}{Security Content Automation Protocol\acroextra{ (by \acs{NIST})}}
    	\acro{SiLK}{System for Internet-level Knowledge\acroextra{, an open-source analysis tool set published by \ac{CERT/CC}}}
	\acro{SMB}{Server Message Block}
    	\acro{SoK}{Systematization of Knowledge\acroextra{ paper in \acs{IEEE} Oakland conference}}
    	\acro{STIX}{Structured Threat Information Expression\acroextra{ (by \acs{MITRE})}}
    	\acro{STS}{Science and Technology Studies\acroextra{ (a field synthesizing philosophy of science, history of science, sociology of science, and philosohpy of technology)}}
    	\acro{TAXII}{Trusted Automated eXchange of Indicator Information\acroextra{ (by \acs{MITRE})}}
    	\acro{TCP/IP}{Transmission Control Protocol / Internet Protocol}
    	\acro{TLA}{Temporal Logic of Actions}
    	\acro{TLP}{Traffic Light Protocol}
    	\acro{TLD}{Top-Level Domain\acroextra{ (in \acs{DNS})}}
	\acro{TLS}{Transport Layer Security}
	\acro{TSA}{Transport Security Administration\acroextra{ (\acs*{US})}}
    	\acro{TTPs}{Tools, tactics, and procedures}
    	\acro{UN}{United Nations}
	\acro{UML}{Unified Modeling Language\acroextra{, see \citet{larman2001patterns}}}
	\acro{US}{United States of America}
    	\acro{USCERT}[US-CERT]{\acs*{US} Computer Emergency Readiness Team\acroextra{, a branch of \acs{NCCIC} within \acs{DHS}}}
	\acro{URL}{Uniform Resource Locator}
	\acro{VERIS}{Vocabulary for Event Recording and Incident Sharing}
    	\acro{W3C}{World Wide Web Consortium}
    	\acro{XCCDF}{Extensible Configuration Checklist Description Format\acroextra{ (by \acs{NIST})}}
    	\acro{XML}{Extensible Markup Language\acroextra{, a standard by \acs{W3C}}}
\end{acronym}

%% file: vuls-in-ML-systems_v1.bbl

\begin{thebibliography}{58}


\ifx \showCODEN    \undefined \def \showCODEN     #1{\unskip}     \fi
\ifx \showDOI      \undefined \def \showDOI       #1{#1}\fi
\ifx \showISBNx    \undefined \def \showISBNx     #1{\unskip}     \fi
\ifx \showISBNxiii \undefined \def \showISBNxiii  #1{\unskip}     \fi
\ifx \showISSN     \undefined \def \showISSN      #1{\unskip}     \fi
\ifx \showLCCN     \undefined \def \showLCCN      #1{\unskip}     \fi
\ifx \shownote     \undefined \def \shownote      #1{#1}          \fi
\ifx \showarticletitle \undefined \def \showarticletitle #1{#1}   \fi
\ifx \showURL      \undefined \def \showURL       {\relax}        \fi
\providecommand\bibfield[2]{#2}
\providecommand\bibinfo[2]{#2}
\providecommand\natexlab[1]{#1}
\providecommand\showeprint[2][]{arXiv:#2}

\bibitem[\protect\citeauthoryear{Becker and Lecun}{Becker and Lecun}{1989}]%
        {becker1989improving}
\bibfield{author}{\bibinfo{person}{S Becker} {and} \bibinfo{person}{Yann
  Lecun}.} \bibinfo{year}{1989}\natexlab{}.
\newblock \showarticletitle{Improving the convergence of back-propagation
  learning with second-order methods}. In \bibinfo{booktitle}{\emph{Proceedings
  of the 1988 Connectionist Models Summer School, San Mateo}}. Morgan Kaufmann,
  \bibinfo{pages}{29--37}.
\newblock


\bibitem[\protect\citeauthoryear{Benetis, Caleff, Hoepers, Horneman,
  Householder, Kossakowski, Manion, Mullens, Perl, Roethlisberger, Rokas,
  Rossell, Ruefle, Sacher, Tzvetanov, and Zajicek}{Benetis
  et~al\mbox{.}}{2019}]%
        {csirtservices_v2}
\bibfield{author}{\bibinfo{person}{Vilius Benetis}, \bibinfo{person}{Olivier
  Caleff}, \bibinfo{person}{Cristine Hoepers}, \bibinfo{person}{Angela
  Horneman}, \bibinfo{person}{Allen Householder}, \bibinfo{person}{Klaus-Peter
  Kossakowski}, \bibinfo{person}{Art Manion}, \bibinfo{person}{Amanda Mullens},
  \bibinfo{person}{Samuel Perl}, \bibinfo{person}{Daniel Roethlisberger},
  \bibinfo{person}{Sigitas Rokas}, \bibinfo{person}{Mary Rossell},
  \bibinfo{person}{Robin~M. Ruefle}, \bibinfo{person}{D{'e}sir{'e}e Sacher},
  \bibinfo{person}{Krassimir~T. Tzvetanov}, {and} \bibinfo{person}{Mark
  Zajicek}.} \bibinfo{year}{2019}\natexlab{}.
\newblock \bibinfo{booktitle}{\emph{Computer Security Incident Response Team
  {(CSIRT)} Services Framework}}.
\newblock \bibinfo{type}{{T}echnical {R}eport} ver. 2.
  \bibinfo{institution}{FIRST}, \bibinfo{address}{Cary, NC, USA}.
\newblock


\bibitem[\protect\citeauthoryear{Biggio and Roli}{Biggio and Roli}{2018}]%
        {biggio2018wild}
\bibfield{author}{\bibinfo{person}{Battista Biggio} {and}
  \bibinfo{person}{Fabio Roli}.} \bibinfo{year}{2018}\natexlab{}.
\newblock \showarticletitle{Wild patterns: Ten years after the rise of
  adversarial machine learning}.
\newblock \bibinfo{journal}{\emph{Pattern Recognition}}  \bibinfo{volume}{84}
  (\bibinfo{year}{2018}), \bibinfo{pages}{317--331}.
\newblock


\bibitem[\protect\citeauthoryear{Buolamwini and Gebru}{Buolamwini and
  Gebru}{2018}]%
        {buolamwini2018gender}
\bibfield{author}{\bibinfo{person}{Joy Buolamwini} {and}
  \bibinfo{person}{Timnit Gebru}.} \bibinfo{year}{2018}\natexlab{}.
\newblock \showarticletitle{Gender shades: Intersectional accuracy disparities
  in commercial gender classification}. In \bibinfo{booktitle}{\emph{Conference
  on fairness, accountability and transparency}}. \bibinfo{pages}{77--91}.
\newblock


\bibitem[\protect\citeauthoryear{Carlini}{Carlini}{2020}]%
        {carlini_nicholas_complete_2020}
\bibfield{author}{\bibinfo{person}{Nicholas Carlini}.}
  \bibinfo{year}{2020}\natexlab{}.
\newblock \bibinfo{title}{A {Complete} {List} of {All} {Adversarial} {Example}
  {Papers}}.
\newblock
\newblock
\urldef\tempurl%
\url{https://nicholas.carlini.com/writing/2019/all-adversarial-example-papers.html}
\showURL{%
\tempurl}


\bibitem[\protect\citeauthoryear{Carlini, Athalye, Papernot, Brendel, Rauber,
  Tsipras, Goodfellow, Madry, and Kurakin}{Carlini et~al\mbox{.}}{2019}]%
        {carlini2019evaluating}
\bibfield{author}{\bibinfo{person}{Nicholas Carlini}, \bibinfo{person}{Anish
  Athalye}, \bibinfo{person}{Nicolas Papernot}, \bibinfo{person}{Wieland
  Brendel}, \bibinfo{person}{Jonas Rauber}, \bibinfo{person}{Dimitris Tsipras},
  \bibinfo{person}{Ian Goodfellow}, \bibinfo{person}{Aleksander Madry}, {and}
  \bibinfo{person}{Alexey Kurakin}.} \bibinfo{year}{2019}\natexlab{}.
\newblock \showarticletitle{On Evaluating Adversarial Robustness}.
\newblock \bibinfo{journal}{\emph{arXiv preprint arXiv:1902.06705}}
  (\bibinfo{year}{2019}).
\newblock


\bibitem[\protect\citeauthoryear{Carlini, Jagielski, and Mironov}{Carlini
  et~al\mbox{.}}{2020}]%
        {carlini_cryptanalytic_2020}
\bibfield{author}{\bibinfo{person}{Nicholas Carlini}, \bibinfo{person}{Matthew
  Jagielski}, {and} \bibinfo{person}{Ilya Mironov}.}
  \bibinfo{year}{2020}\natexlab{}.
\newblock \showarticletitle{Cryptanalytic {Extraction} of {Neural} {Network}
  {Models}}.
\newblock \bibinfo{journal}{\emph{arXiv:2003.04884 [cs]}} (\bibinfo{date}{July}
  \bibinfo{year}{2020}).
\newblock
\urldef\tempurl%
\url{http://arxiv.org/abs/2003.04884}
\showURL{%
\tempurl}
\newblock
\shownote{arXiv: 2003.04884.}


\bibitem[\protect\citeauthoryear{Citron and Pasquale}{Citron and
  Pasquale}{2014}]%
        {Citron14}
\bibfield{author}{\bibinfo{person}{Danielle~Keats Citron} {and}
  \bibinfo{person}{Frank~A. Pasquale}.} \bibinfo{year}{2014}\natexlab{}.
\newblock \showarticletitle{The Scored Society: Due Process for Automated
  Predictions}.
\newblock \bibinfo{journal}{\emph{Washington Law Review}} \bibinfo{volume}{89},
  \bibinfo{number}{8} (\bibinfo{year}{2014}).
\newblock
\urldef\tempurl%
\url{https://ssrn.com/abstract=2376209}
\showURL{%
\tempurl}


\bibitem[\protect\citeauthoryear{{CVE Board}}{{CVE Board}}{2020}]%
        {mitre2020cna}
\bibfield{author}{\bibinfo{person}{{CVE Board}}.}
  \bibinfo{year}{2020}\natexlab{}.
\newblock \bibinfo{booktitle}{\emph{{CVE} Numbering Authority {(CNA)} rules}}.
\newblock \bibinfo{type}{{T}echnical {R}eport} ver. 3.0.
  \bibinfo{institution}{MITRE}, \bibinfo{address}{Bedford, MA}.
\newblock
\urldef\tempurl%
\url{https://cve.mitre.org/cve/cna/rules.html}
\showURL{%
\tempurl}


\bibitem[\protect\citeauthoryear{{CVSS SIG}}{{CVSS SIG}}{2019}]%
        {cvss_v3-1}
\bibfield{author}{\bibinfo{person}{{CVSS SIG}}.}
  \bibinfo{year}{2019}\natexlab{}.
\newblock \bibinfo{booktitle}{\emph{Common Vulnerability Scoring System}}.
\newblock \bibinfo{type}{{T}echnical {R}eport} version 3.1 r1.
  \bibinfo{institution}{Forum of Incident Response and Security Teams},
  \bibinfo{address}{Cary, NC, USA}.
\newblock
\urldef\tempurl%
\url{https://www.first.org/cvss/v3.1/specification-document}
\showURL{%
\tempurl}


\bibitem[\protect\citeauthoryear{Dougherty}{Dougherty}{2008}]%
        {vu836068MD5}
\bibfield{author}{\bibinfo{person}{Chad~R Dougherty}.}
  \bibinfo{year}{2008}\natexlab{}.
\newblock \bibinfo{title}{VU\#836068: MD5 vulnerable to collision attacks}.
\newblock
\newblock
\urldef\tempurl%
\url{https://kb.cert.org/vuls/id/836068}
\showURL{%
\tempurl}
\newblock
\shownote{Accessed 2020-08-10.}


\bibitem[\protect\citeauthoryear{Eidelman}{Eidelman}{2018}]%
        {Eidelman18}
\bibfield{author}{\bibinfo{person}{Vera Eidelman}.}
  \bibinfo{year}{2018}\natexlab{}.
\newblock \showarticletitle{The First Amendment Case for Public Access to
  Secret Algorithms Used in Criminal Trials}.
\newblock \bibinfo{journal}{\emph{Georgia State University Law Review}}
  \bibinfo{volume}{34}, \bibinfo{number}{4} (\bibinfo{date}{August}
  \bibinfo{year}{2018}).
\newblock


\bibitem[\protect\citeauthoryear{Farnadi, Babaki, and Getoor}{Farnadi
  et~al\mbox{.}}{2018}]%
        {Farnadi18}
\bibfield{author}{\bibinfo{person}{Golnoosh Farnadi}, \bibinfo{person}{Behrouz
  Babaki}, {and} \bibinfo{person}{Lise Getoor}.}
  \bibinfo{year}{2018}\natexlab{}.
\newblock \showarticletitle{Fairness in Relational Domains}. In
  \bibinfo{booktitle}{\emph{AIES '18: Proceedings of the 2018 AAAI/ACM
  Conference on AI, Ethics, and Society}},
  \bibfield{editor}{\bibinfo{person}{Jason Furman}, \bibinfo{person}{Gary
  Marchant}, \bibinfo{person}{Huw Price}, {and} \bibinfo{person}{Francesca
  Rossi}} (Eds.). \bibinfo{address}{New Orleans, LA, USA}.
\newblock
\urldef\tempurl%
\url{https://doi.org/10.1145/3278721.3278733}
\showDOI{\tempurl}


\bibitem[\protect\citeauthoryear{Fox, Arnoth, Skorupka, Mc{C}ollum, and
  Bodeau}{Fox et~al\mbox{.}}{2018}]%
        {fox2018threat}
\bibfield{author}{\bibinfo{person}{David Fox}, \bibinfo{person}{Eric Arnoth},
  \bibinfo{person}{Clement Skorupka}, \bibinfo{person}{Catherine Mc{C}ollum},
  {and} \bibinfo{person}{Deborah Bodeau}.} \bibinfo{year}{2018}\natexlab{}.
\newblock \bibinfo{booktitle}{\emph{Enhanced Cyber Threat Model for Financial
  Services Sector (FSS) Institutions}}.
\newblock \bibinfo{type}{{T}echnical {R}eport} 18-1725.
  \bibinfo{institution}{Homeland Security Systems Engineering and Development
  Institute}, \bibinfo{address}{McLean, VA}.
\newblock


\bibitem[\protect\citeauthoryear{Galison}{Galison}{1999}]%
        {galison1999trading}
\bibfield{author}{\bibinfo{person}{Peter Galison}.}
  \bibinfo{year}{1999}\natexlab{}.
\newblock \showarticletitle{Trading zone: Coordinating action and belief}.
\newblock \bibinfo{journal}{\emph{The Science Studies Reader}}
  (\bibinfo{year}{1999}), \bibinfo{pages}{137--160}.
\newblock


\bibitem[\protect\citeauthoryear{Galyardt, VanHoudnos, and Spring}{Galyardt
  et~al\mbox{.}}{2020}]%
        {galyardt2020comments}
\bibfield{author}{\bibinfo{person}{April Galyardt}, \bibinfo{person}{Nathan~M.
  VanHoudnos}, {and} \bibinfo{person}{Jonathan~M. Spring}.}
  \bibinfo{year}{2020}\natexlab{}.
\newblock \bibinfo{booktitle}{\emph{Comments on {NISTIR} 8269 (A Taxonomy and
  Terminology of Adversarial Machine Learning)}}.
\newblock \bibinfo{type}{{T}echnical {R}eport}. \bibinfo{institution}{Software
  Engineering Institute, Carnegie Mellon University},
  \bibinfo{address}{Pittsburgh, PA}.
\newblock
\urldef\tempurl%
\url{https://resources.sei.cmu.edu/library/asset-view.cfm?assetid=637327}
\showURL{%
\tempurl}


\bibitem[\protect\citeauthoryear{Grier, Ballard, Caballero, Chachra, Dietrich,
  Levchenko, Mavrommatis, McCoy, Nappa, Pitsillidis, Provos, Rafique, Rajab,
  Rossow, Thomas, Paxson, Savage, and Voelker}{Grier et~al\mbox{.}}{2012}]%
        {Grier:2012:MCE:2382196.2382283}
\bibfield{author}{\bibinfo{person}{Chris Grier}, \bibinfo{person}{Lucas
  Ballard}, \bibinfo{person}{Juan Caballero}, \bibinfo{person}{Neha Chachra},
  \bibinfo{person}{Christian~J. Dietrich}, \bibinfo{person}{Kirill Levchenko},
  \bibinfo{person}{Panayiotis Mavrommatis}, \bibinfo{person}{Damon McCoy},
  \bibinfo{person}{Antonio Nappa}, \bibinfo{person}{Andreas Pitsillidis},
  \bibinfo{person}{Niels Provos}, \bibinfo{person}{M.~Zubair Rafique},
  \bibinfo{person}{Moheeb~Abu Rajab}, \bibinfo{person}{Christian Rossow},
  \bibinfo{person}{Kurt Thomas}, \bibinfo{person}{Vern Paxson},
  \bibinfo{person}{Stefan Savage}, {and} \bibinfo{person}{Geoffrey~M.
  Voelker}.} \bibinfo{year}{2012}\natexlab{}.
\newblock \showarticletitle{Manufacturing Compromise: The Emergence of
  Exploit-as-a-service}. In \bibinfo{booktitle}{\emph{Conference on Computer
  and Communications Security}}. \bibinfo{publisher}{ACM},
  \bibinfo{address}{Raleigh, North Carolina, USA}, \bibinfo{pages}{821--832}.
\newblock


\bibitem[\protect\citeauthoryear{Gu, Dolan-Gavitt, and Garg}{Gu
  et~al\mbox{.}}{2019}]%
        {gu_badnets_2019}
\bibfield{author}{\bibinfo{person}{Tianyu Gu}, \bibinfo{person}{Brendan
  Dolan-Gavitt}, {and} \bibinfo{person}{Siddharth Garg}.}
  \bibinfo{year}{2019}\natexlab{}.
\newblock \showarticletitle{{BadNets}: {Identifying} {Vulnerabilities} in the
  {Machine} {Learning} {Model} {Supply} {Chain}}.
\newblock \bibinfo{journal}{\emph{arXiv:1708.06733 [cs]}}
  (\bibinfo{date}{March} \bibinfo{year}{2019}).
\newblock
\urldef\tempurl%
\url{http://arxiv.org/abs/1708.06733}
\showURL{%
\tempurl}
\newblock
\shownote{arXiv: 1708.06733.}


\bibitem[\protect\citeauthoryear{{Health ISAC}}{{Health ISAC}}{2019}]%
        {hisacCVD}
\bibfield{author}{\bibinfo{person}{{Health ISAC}}.}
  \bibinfo{year}{2019}\natexlab{}.
\newblock \bibinfo{title}{Medical Device Security Media Education Materials}.
\newblock
\newblock
\urldef\tempurl%
\url{https://h-isac.org/cvd-media-kit/}
\showURL{%
\tempurl}
\newblock
\shownote{Accessed Aug 18, 2020.}


\bibitem[\protect\citeauthoryear{Horneman, Mellinger, and Ozkaya}{Horneman
  et~al\mbox{.}}{2019}]%
        {horneman_ai_2019}
\bibfield{author}{\bibinfo{person}{Angela Horneman}, \bibinfo{person}{Andrew
  Mellinger}, {and} \bibinfo{person}{Ipek Ozkaya}.}
  \bibinfo{year}{2019}\natexlab{}.
\newblock \bibinfo{booktitle}{\emph{{AI} {Engineering}: 11 {Foundational}
  {Practices}}}.
\newblock \bibinfo{type}{{T}echnical {R}eport}. \bibinfo{institution}{Software
  Engineering Institute, Carnegie Mellon University},
  \bibinfo{address}{Pittsburgh, PA}.
\newblock


\bibitem[\protect\citeauthoryear{Householder}{Householder}{2015}]%
        {Householder2015steampunk}
\bibfield{author}{\bibinfo{person}{Allen Householder}.}
  \bibinfo{year}{2015}\natexlab{}.
\newblock \bibinfo{booktitle}{\emph{Systemic Vulnerabilities: An Allegorical
  Tale of Steampunk Vulnerability to Aero-Physical Threats}}.
\newblock CERT Coordination Center, Software Engineering Institute, Carnegie
  Mellon University.
\newblock
\urldef\tempurl%
\url{https://www.youtube.com/watch?v=4AHpL3kVHw4}
\showURL{%
\tempurl}


\bibitem[\protect\citeauthoryear{Householder, Spring, VanHoudnos, and
  Wright}{Householder et~al\mbox{.}}{2020a}]%
        {vu425163}
\bibfield{author}{\bibinfo{person}{Allen Householder},
  \bibinfo{person}{Jonathan~M. Spring}, \bibinfo{person}{Nathan VanHoudnos},
  {and} \bibinfo{person}{Oren Wright}.} \bibinfo{year}{2020}\natexlab{a}.
\newblock \bibinfo{title}{Machine learning classifiers trained via gradient
  descent are vulnerable to arbitrary misclassification attack}.
\newblock
\newblock
\urldef\tempurl%
\url{https://kb.cert.org/vuls/id/425163/}
\showURL{%
\tempurl}


\bibitem[\protect\citeauthoryear{Householder, Wassermann, Manion, and
  King}{Householder et~al\mbox{.}}{2020b}]%
        {householder2020cvd}
\bibfield{author}{\bibinfo{person}{Allen~D. Householder},
  \bibinfo{person}{Garret Wassermann}, \bibinfo{person}{Art Manion}, {and}
  \bibinfo{person}{Christopher King}.} \bibinfo{year}{2020}\natexlab{b}.
\newblock \bibinfo{booktitle}{\emph{The {CERT}\textregistered ~Guide to
  Coordinated Vulnerability Disclosure}}.
\newblock \bibinfo{type}{{T}echnical {R}eport} CMU/SEI-2017-TR-022.
  \bibinfo{institution}{Software Engineering Institute, Carnegie Mellon
  University}, \bibinfo{address}{Pittsburgh, PA}.
\newblock
\urldef\tempurl%
\url{https://vuls.cert.org/confluence/display/CVD}
\showURL{%
\tempurl}


\bibitem[\protect\citeauthoryear{Jump and Manion}{Jump and Manion}{2019}]%
        {manion2019sbom}
\bibfield{author}{\bibinfo{person}{Michelle Jump} {and} \bibinfo{person}{Art
  Manion}.} \bibinfo{year}{2019}\natexlab{}.
\newblock \bibinfo{booktitle}{\emph{Framing Software Component Transparency:
  {E}stablishing a Common Software Bill of Material ({SBOM})}}.
\newblock \bibinfo{type}{{T}echnical {R}eport}. \bibinfo{institution}{National
  Telecommunications and Information Administration},
  \bibinfo{address}{Washington, DC}.
\newblock


\bibitem[\protect\citeauthoryear{Kroll, Huey, Barocas, Felten, Reidenberg,
  Robinson, and Yu}{Kroll et~al\mbox{.}}{2017}]%
        {Kroll17}
\bibfield{author}{\bibinfo{person}{Joshua~A. Kroll}, \bibinfo{person}{Joanna
  Huey}, \bibinfo{person}{Solon Barocas}, \bibinfo{person}{Edward~W. Felten},
  \bibinfo{person}{Joel~R. Reidenberg}, \bibinfo{person}{David~G. Robinson},
  {and} \bibinfo{person}{Harlan Yu}.} \bibinfo{year}{2017}\natexlab{}.
\newblock \showarticletitle{Accountable Algorithms}.
\newblock \bibinfo{journal}{\emph{U. PA. Law Review}} (\bibinfo{year}{2017}),
  \bibinfo{pages}{633--}.
\newblock
\urldef\tempurl%
\url{https://scholarship.law.upenn.edu/penn_law_review/vol165/iss3/3}
\showURL{%
\tempurl}


\bibitem[\protect\citeauthoryear{Kumar, Nyström, Lambert, Marshall, Goertzel,
  Comissoneru, Swann, and Xia}{Kumar et~al\mbox{.}}{2020}]%
        {kumar2020adversarial}
\bibfield{author}{\bibinfo{person}{Ram Shankar~Siva Kumar},
  \bibinfo{person}{Magnus Nyström}, \bibinfo{person}{John Lambert},
  \bibinfo{person}{Andrew Marshall}, \bibinfo{person}{Mario Goertzel},
  \bibinfo{person}{Andi Comissoneru}, \bibinfo{person}{Matt Swann}, {and}
  \bibinfo{person}{Sharon Xia}.} \bibinfo{year}{2020}\natexlab{}.
\newblock \bibinfo{title}{Adversarial Machine Learning -- Industry
  Perspectives}.
\newblock
\newblock
\showeprint[arxiv]{cs.CY/2002.05646}


\bibitem[\protect\citeauthoryear{Kumar, O'{B}rien, Albert, and Viljoen}{Kumar
  et~al\mbox{.}}{2018}]%
        {kumar2018law}
\bibfield{author}{\bibinfo{person}{Ram Shankar~Siva Kumar},
  \bibinfo{person}{David~R. O'{B}rien}, \bibinfo{person}{Kendra Albert}, {and}
  \bibinfo{person}{Salomé Viljoen}.} \bibinfo{year}{2018}\natexlab{}.
\newblock \bibinfo{title}{Law and Adversarial Machine Learning}.
\newblock
\newblock
\showeprint[arxiv]{cs.LG/1810.10731}


\bibitem[\protect\citeauthoryear{Kumar, O'{B}rien, Albert, Viljoen, and
  Snover}{Kumar et~al\mbox{.}}{2019}]%
        {kumar2019failure}
\bibfield{author}{\bibinfo{person}{Ram Shankar~Siva Kumar},
  \bibinfo{person}{David~R. O'{B}rien}, \bibinfo{person}{Kendra Albert},
  \bibinfo{person}{Salomé Viljoen}, {and} \bibinfo{person}{Jeffrey Snover}.}
  \bibinfo{year}{2019}\natexlab{}.
\newblock \showarticletitle{Failure Modes in Machine Learning Systems}.
\newblock \bibinfo{journal}{\emph{arXiv preprint}} \bibinfo{number}{1911.11034}
  (\bibinfo{year}{2019}).
\newblock


\bibitem[\protect\citeauthoryear{Lehr and Ohm}{Lehr and Ohm}{2017}]%
        {Lehr17}
\bibfield{author}{\bibinfo{person}{David Lehr} {and} \bibinfo{person}{Paul
  Ohm}.} \bibinfo{year}{2017}\natexlab{}.
\newblock \showarticletitle{Playing with the Data: What Legal Scholars Should
  Learn About Machine Learning}.
\newblock \bibinfo{journal}{\emph{UCDL Rev.}}  \bibinfo{volume}{51}
  (\bibinfo{year}{2017}), \bibinfo{pages}{653}.
\newblock


\bibitem[\protect\citeauthoryear{Lin, Maire, Belongie, Bourdev, Girshick, Hays,
  Perona, Ramanan, Zitnick, and Dollár}{Lin et~al\mbox{.}}{2015}]%
        {lin_microsoft_2015}
\bibfield{author}{\bibinfo{person}{Tsung-Yi Lin}, \bibinfo{person}{Michael
  Maire}, \bibinfo{person}{Serge Belongie}, \bibinfo{person}{Lubomir Bourdev},
  \bibinfo{person}{Ross Girshick}, \bibinfo{person}{James Hays},
  \bibinfo{person}{Pietro Perona}, \bibinfo{person}{Deva Ramanan},
  \bibinfo{person}{C.~Lawrence Zitnick}, {and} \bibinfo{person}{Piotr
  Dollár}.} \bibinfo{year}{2015}\natexlab{}.
\newblock \showarticletitle{Microsoft {COCO}: {Common} {Objects} in {Context}}.
\newblock \bibinfo{journal}{\emph{arXiv:1405.0312 [cs]}} (\bibinfo{date}{Feb.}
  \bibinfo{year}{2015}).
\newblock
\urldef\tempurl%
\url{http://arxiv.org/abs/1405.0312}
\showURL{%
\tempurl}
\newblock
\shownote{arXiv: 1405.0312.}


\bibitem[\protect\citeauthoryear{Man{\`e}s, Han, Han, Cha, Egele, Schwartz, and
  Woo}{Man{\`e}s et~al\mbox{.}}{2019}]%
        {manes2019art}
\bibfield{author}{\bibinfo{person}{Valentin Jean~Marie Man{\`e}s},
  \bibinfo{person}{HyungSeok Han}, \bibinfo{person}{Choongwoo Han},
  \bibinfo{person}{Sang~Kil Cha}, \bibinfo{person}{Manuel Egele},
  \bibinfo{person}{Edward~J Schwartz}, {and} \bibinfo{person}{Maverick Woo}.}
  \bibinfo{year}{2019}\natexlab{}.
\newblock \showarticletitle{The art, science, and engineering of fuzzing: A
  survey}.
\newblock \bibinfo{journal}{\emph{IEEE Transactions on Software Engineering}}
  (\bibinfo{year}{2019}).
\newblock


\bibitem[\protect\citeauthoryear{Matsui}{Matsui}{1993}]%
        {matsui1993des}
\bibfield{author}{\bibinfo{person}{Mitsuru Matsui}.}
  \bibinfo{year}{1993}\natexlab{}.
\newblock \showarticletitle{Linear Cryptanalysis Method for DES Cipher}. In
  \bibinfo{booktitle}{\emph{Advances in Cryptology --- EUROCRYPT}}
  \emph{(\bibinfo{series}{LNCS 765})}, \bibfield{editor}{\bibinfo{person}{Tor
  Helleseth}} (Ed.). \bibinfo{publisher}{Springer}, \bibinfo{address}{Lofthus,
  Norway}, \bibinfo{pages}{386--397}.
\newblock


\bibitem[\protect\citeauthoryear{{MITRE Corporation}}{{MITRE
  Corporation}}{2010}]%
        {mitre_cve}
\bibfield{author}{\bibinfo{person}{{MITRE Corporation}}.}
  \bibinfo{year}{2010}\natexlab{}.
\newblock \bibinfo{title}{Common Vulnerabilities and Exposures}.
\newblock \bibinfo{howpublished}{\url{http://cve.mitre.org}}.
\newblock
\newblock
\shownote{last access May 2, 2020.}


\bibitem[\protect\citeauthoryear{{MITRE Corporation}}{{MITRE
  Corporation}}{2014}]%
        {mitre_cwe}
\bibfield{author}{\bibinfo{person}{{MITRE Corporation}}.}
  \bibinfo{year}{2014}\natexlab{}.
\newblock \bibinfo{title}{Common Weakness Enumeration: A Community-Developed
  Dictionary of Software Weakness Types}.
\newblock \bibinfo{howpublished}{\url{http://cwe.mitre.org}}.
\newblock


\bibitem[\protect\citeauthoryear{Noble}{Noble}{2018}]%
        {noble2018algorithms}
\bibfield{author}{\bibinfo{person}{Safiya~Umoja Noble}.}
  \bibinfo{year}{2018}\natexlab{}.
\newblock \bibinfo{booktitle}{\emph{Algorithms of oppression: How search
  engines reinforce racism}}.
\newblock \bibinfo{publisher}{NYU Press}, \bibinfo{address}{New York, NY}.
\newblock


\bibitem[\protect\citeauthoryear{O'{N}eil}{O'{N}eil}{2016}]%
        {oneil2016weapons}
\bibfield{author}{\bibinfo{person}{Cathy O'{N}eil}.}
  \bibinfo{year}{2016}\natexlab{}.
\newblock \bibinfo{booktitle}{\emph{Weapons of math destruction: How big data
  increases inequality and threatens democracy}}.
\newblock \bibinfo{publisher}{Broadway Books}, \bibinfo{address}{New York, NY}.
\newblock


\bibitem[\protect\citeauthoryear{{OWASP Foundation}}{{OWASP
  Foundation}}{2017}]%
        {owasptop10}
\bibfield{author}{\bibinfo{person}{{OWASP Foundation}}.}
  \bibinfo{year}{2017}\natexlab{}.
\newblock \bibinfo{title}{OWASP Top Ten}.
\newblock
\newblock
\urldef\tempurl%
\url{https://owasp.org/www-project-top-ten/}
\showURL{%
\tempurl}
\newblock
\shownote{Accessed 2020-08-10.}


\bibitem[\protect\citeauthoryear{Papernot, Faghri, Carlini, Goodfellow,
  Feinman, Kurakin, Xie, Sharma, Brown, Roy, Matyasko, Behzadan, Hambardzumyan,
  Zhang, Juang, Li, Sheatsley, Garg, Uesato, Gierke, Dong, Berthelot,
  Hendricks, Rauber, Long, and McDaniel}{Papernot et~al\mbox{.}}{2016}]%
        {papernot2016technical}
\bibfield{author}{\bibinfo{person}{Nicolas Papernot}, \bibinfo{person}{Fartash
  Faghri}, \bibinfo{person}{Nicholas Carlini}, \bibinfo{person}{Ian
  Goodfellow}, \bibinfo{person}{Reuben Feinman}, \bibinfo{person}{Alexey
  Kurakin}, \bibinfo{person}{Cihang Xie}, \bibinfo{person}{Yash Sharma},
  \bibinfo{person}{Tom Brown}, \bibinfo{person}{Aurko Roy},
  \bibinfo{person}{Alexander Matyasko}, \bibinfo{person}{Vahid Behzadan},
  \bibinfo{person}{Karen Hambardzumyan}, \bibinfo{person}{Zhishuai Zhang},
  \bibinfo{person}{Yi-Lin Juang}, \bibinfo{person}{Zhi Li},
  \bibinfo{person}{Ryan Sheatsley}, \bibinfo{person}{Abhibhav Garg},
  \bibinfo{person}{Jonathan Uesato}, \bibinfo{person}{Willi Gierke},
  \bibinfo{person}{Yinpeng Dong}, \bibinfo{person}{David Berthelot},
  \bibinfo{person}{Paul Hendricks}, \bibinfo{person}{Jonas Rauber},
  \bibinfo{person}{Rujun Long}, {and} \bibinfo{person}{Patrick McDaniel}.}
  \bibinfo{year}{2016}\natexlab{}.
\newblock \showarticletitle{Technical Report on the CleverHans v2.1.0
  Adversarial Examples Library}.
\newblock \bibinfo{journal}{\emph{arXiv}} \bibinfo{number}{1610.00768}
  (\bibinfo{year}{2016}).
\newblock


\bibitem[\protect\citeauthoryear{Papernot, McDaniel, Sinha, and
  Wellman}{Papernot et~al\mbox{.}}{2018}]%
        {papernot2018sok}
\bibfield{author}{\bibinfo{person}{Nicolas Papernot}, \bibinfo{person}{Patrick
  McDaniel}, \bibinfo{person}{Arunesh Sinha}, {and} \bibinfo{person}{Michael~P
  Wellman}.} \bibinfo{year}{2018}\natexlab{}.
\newblock \showarticletitle{SoK: {S}ecurity and privacy in machine learning}.
  In \bibinfo{booktitle}{\emph{European Symposium on Security and Privacy}}.
  \bibinfo{publisher}{IEEE}, \bibinfo{address}{London, UK},
  \bibinfo{pages}{399--414}.
\newblock


\bibitem[\protect\citeauthoryear{Pieters}{Pieters}{2011}]%
        {pieters2011social}
\bibfield{author}{\bibinfo{person}{Wolter Pieters}.}
  \bibinfo{year}{2011}\natexlab{}.
\newblock \showarticletitle{The (social) construction of information security}.
\newblock \bibinfo{journal}{\emph{The Information Society}}
  \bibinfo{volume}{27}, \bibinfo{number}{5} (\bibinfo{year}{2011}),
  \bibinfo{pages}{326--335}.
\newblock


\bibitem[\protect\citeauthoryear{Rawls and Mann}{Rawls and Mann}{2010}]%
        {rawls2010thing}
\bibfield{author}{\bibinfo{person}{Anne~W Rawls} {and} \bibinfo{person}{David
  Mann}.} \bibinfo{year}{2010}\natexlab{}.
\newblock \bibinfo{booktitle}{\emph{The Thing is What is Our `What': An
  Ethnographic Study of a Design Team's Discussion of `Object' Clarity as a
  Problem in Designing an Information System to Facilitate System
  Interoperability}}.
\newblock \bibinfo{type}{{T}echnical {R}eport} 10-2594.
  \bibinfo{institution}{MITRE Corp}, \bibinfo{address}{McLean, VA, United
  States}.
\newblock


\bibitem[\protect\citeauthoryear{Redmon}{Redmon}{2016}]%
        {yolov2_weights}
\bibfield{author}{\bibinfo{person}{Joseph Redmon}.}
  \bibinfo{year}{2016}\natexlab{}.
\newblock \bibinfo{title}{yolov2.weights}.
\newblock
\newblock
\urldef\tempurl%
\url{https://pjreddie.com/media/files/yolov2.weights}
\showURL{%
\tempurl}
\newblock
\shownote{Accessed Aug 10, 2020.}


\bibitem[\protect\citeauthoryear{Redmon and Farhadi}{Redmon and
  Farhadi}{2016}]%
        {redmon2016yolo9000}
\bibfield{author}{\bibinfo{person}{Joseph Redmon} {and} \bibinfo{person}{Ali
  Farhadi}.} \bibinfo{year}{2016}\natexlab{}.
\newblock \showarticletitle{YOLO9000: Better, Faster, Stronger}.
\newblock \bibinfo{journal}{\emph{arXiv preprint arXiv:1612.08242}}
  (\bibinfo{year}{2016}).
\newblock


\bibitem[\protect\citeauthoryear{Ren, He, Girshick, and Sun}{Ren
  et~al\mbox{.}}{2016}]%
        {ren_faster_2016}
\bibfield{author}{\bibinfo{person}{Shaoqing Ren}, \bibinfo{person}{Kaiming He},
  \bibinfo{person}{Ross Girshick}, {and} \bibinfo{person}{Jian Sun}.}
  \bibinfo{year}{2016}\natexlab{}.
\newblock \showarticletitle{Faster {R}-{CNN}: {Towards} {Real}-{Time} {Object}
  {Detection} with {Region} {Proposal} {Networks}}.
\newblock \bibinfo{journal}{\emph{arXiv:1506.01497 [cs]}} (\bibinfo{date}{Jan.}
  \bibinfo{year}{2016}).
\newblock
\urldef\tempurl%
\url{http://arxiv.org/abs/1506.01497}
\showURL{%
\tempurl}
\newblock
\shownote{arXiv: 1506.01497.}


\bibitem[\protect\citeauthoryear{Rivest}{Rivest}{1992}]%
        {rfc1321}
\bibfield{author}{\bibinfo{person}{R. Rivest}.}
  \bibinfo{year}{1992}\natexlab{}.
\newblock \bibinfo{title}{{The MD5 Message-Digest Algorithm}}.
\newblock \bibinfo{howpublished}{RFC 1321 (Informational)}. ,
  \bibinfo{numpages}{21}~pages.
\newblock
\showISSN{2070-1721}
\urldef\tempurl%
\url{https://www.rfc-editor.org/rfc/rfc1321.txt}
\showURL{%
\tempurl}
\newblock
\shownote{Updated by RFC 6151.}


\bibitem[\protect\citeauthoryear{Ruder}{Ruder}{2016}]%
        {ruder2016overview}
\bibfield{author}{\bibinfo{person}{Sebastian Ruder}.}
  \bibinfo{year}{2016}\natexlab{}.
\newblock \showarticletitle{An overview of gradient descent optimization
  algorithms}.
\newblock \bibinfo{journal}{\emph{arXiv preprint arXiv:1609.04747}}
  (\bibinfo{year}{2016}).
\newblock


\bibitem[\protect\citeauthoryear{Russell, Dewey, and Tegmark}{Russell
  et~al\mbox{.}}{2015}]%
        {Russell15}
\bibfield{author}{\bibinfo{person}{Stuart Russell}, \bibinfo{person}{Daniel
  Dewey}, {and} \bibinfo{person}{Max Tegmark}.}
  \bibinfo{year}{2015}\natexlab{}.
\newblock \showarticletitle{Research Priorities for Robust and Beneficiall
  Artificial Inteligence}.
\newblock \bibinfo{journal}{\emph{AI Magazine}} \bibinfo{volume}{36},
  \bibinfo{number}{4} (\bibinfo{year}{2015}).
\newblock


\bibitem[\protect\citeauthoryear{Seacord}{Seacord}{2005}]%
        {seacord2005secure}
\bibfield{author}{\bibinfo{person}{Robert~C Seacord}.}
  \bibinfo{year}{2005}\natexlab{}.
\newblock \bibinfo{booktitle}{\emph{Secure Coding in C and C++}}.
\newblock \bibinfo{publisher}{Pearson Education}, \bibinfo{address}{Upper
  Saddle Ridge, NJ}.
\newblock


\bibitem[\protect\citeauthoryear{Shirey}{Shirey}{2007}]%
        {rfc4949}
\bibfield{author}{\bibinfo{person}{R. Shirey}.}
  \bibinfo{year}{2007}\natexlab{}.
\newblock \bibinfo{title}{{Internet Security Glossary, Version 2}}.
\newblock \bibinfo{howpublished}{RFC 4949 (Informational)}.
\newblock
\showISSN{2070-1721}


\bibitem[\protect\citeauthoryear{Sood and Enbody}{Sood and Enbody}{2013}]%
        {sood2013crimeware}
\bibfield{author}{\bibinfo{person}{Aditya~K Sood} {and}
  \bibinfo{person}{Richard~J Enbody}.} \bibinfo{year}{2013}\natexlab{}.
\newblock \showarticletitle{Crimeware-as-a-service: A survey of commoditized
  crimeware in the underground market}.
\newblock \bibinfo{journal}{\emph{International Journal of Critical
  Infrastructure Protection}} \bibinfo{volume}{6}, \bibinfo{number}{1}
  (\bibinfo{year}{2013}), \bibinfo{pages}{28--38}.
\newblock


\bibitem[\protect\citeauthoryear{Spring, Fallon, Galyardt, Horneman, Metcalf,
  and Stoner}{Spring et~al\mbox{.}}{2019}]%
        {spring2019ml}
\bibfield{author}{\bibinfo{person}{Jonathan~M. Spring}, \bibinfo{person}{Joshua
  Fallon}, \bibinfo{person}{April Galyardt}, \bibinfo{person}{Angela Horneman},
  \bibinfo{person}{Leigh Metcalf}, {and} \bibinfo{person}{Ed Stoner}.}
  \bibinfo{year}{2019}\natexlab{}.
\newblock \bibinfo{booktitle}{\emph{Machine Learning in Cybersecurity: A
  Guide}}.
\newblock \bibinfo{type}{{T}echnical {R}eport} CMU/SEI-2019-TR-005.
  \bibinfo{institution}{Software Engineering Institute, Carnegie Mellon
  University}, \bibinfo{address}{Pittsburgh, PA}.
\newblock
\urldef\tempurl%
\url{http://resources.sei.cmu.edu/library/asset-view.cfm?AssetID=633583}
\showURL{%
\tempurl}


\bibitem[\protect\citeauthoryear{Spring, Hatleback, Householder, Manion, and
  Shick}{Spring et~al\mbox{.}}{2020}]%
        {spring2020ssvc}
\bibfield{author}{\bibinfo{person}{Jonathan~M Spring}, \bibinfo{person}{Eric
  Hatleback}, \bibinfo{person}{Allen~D. Householder}, \bibinfo{person}{Art
  Manion}, {and} \bibinfo{person}{Deana Shick}.}
  \bibinfo{year}{2020}\natexlab{}.
\newblock \showarticletitle{Prioritizing vulnerability response: {A}
  stakeholder-specific vulnerability categorization}. In
  \bibinfo{booktitle}{\emph{Workshop on the Economics of Information
  Security}}. \bibinfo{address}{Brussels, Belgium}.
\newblock


\bibitem[\protect\citeauthoryear{Spring, Moore, and Pym}{Spring
  et~al\mbox{.}}{2017}]%
        {spring2017why}
\bibfield{author}{\bibinfo{person}{Jonathan~M Spring}, \bibinfo{person}{Tyler
  Moore}, {and} \bibinfo{person}{David Pym}.} \bibinfo{year}{2017}\natexlab{}.
\newblock \showarticletitle{Practicing a Science of Security: A philosophy of
  science perspective}. In \bibinfo{booktitle}{\emph{New Security Paradigms
  Workshop}}. \bibinfo{publisher}{ACM}, \bibinfo{address}{Santa Cruz, CA, USA}.
\newblock


\bibitem[\protect\citeauthoryear{Tabassi, Burns, Hadjimichael, Molina-Markham,
  and Sexton}{Tabassi et~al\mbox{.}}{2019}]%
        {nistir8269}
\bibfield{author}{\bibinfo{person}{Elham Tabassi}, \bibinfo{person}{Kevin
  Burns}, \bibinfo{person}{Michael Hadjimichael}, \bibinfo{person}{Andres
  Molina-Markham}, {and} \bibinfo{person}{Julian Sexton}.}
  \bibinfo{year}{2019}\natexlab{}.
\newblock \bibinfo{booktitle}{\emph{A Taxonomy and Terminology of Adversarial
  Machine Learning}}.
\newblock \bibinfo{type}{{T}echnical {R}eport} Draft NISTIR 8269.
  \bibinfo{institution}{National Institute of Standards and Technology},
  \bibinfo{address}{Gathersburg, MD, USA}.
\newblock
\urldef\tempurl%
\url{https://csrc.nist.gov/publications/detail/nistir/8269/draft}
\showURL{%
\tempurl}


\bibitem[\protect\citeauthoryear{{Torchvision}}{{Torchvision}}{2017}]%
        {torchvision_fasterrcnn_resnet50_fpn_coco}
\bibfield{author}{\bibinfo{person}{{Torchvision}}.}
  \bibinfo{year}{2017}\natexlab{}.
\newblock \bibinfo{title}{fasterrcnn\_resnet50\_fpn\_coco}.
\newblock
\newblock
\urldef\tempurl%
\url{https://download.pytorch.org/models/fasterrcnn_resnet50_fpn_coco-258fb6c6.pth}
\showURL{%
\tempurl}
\newblock
\shownote{Accessed Aug 10, 2020.}


\bibitem[\protect\citeauthoryear{Tramer, Carlini, Brendel, and Madry}{Tramer
  et~al\mbox{.}}{2020}]%
        {tramer_adaptive_2020}
\bibfield{author}{\bibinfo{person}{Florian Tramer}, \bibinfo{person}{Nicholas
  Carlini}, \bibinfo{person}{Wieland Brendel}, {and}
  \bibinfo{person}{Aleksander Madry}.} \bibinfo{year}{2020}\natexlab{}.
\newblock \showarticletitle{On {Adaptive} {Attacks} to {Adversarial} {Example}
  {Defenses}}.
\newblock \bibinfo{journal}{\emph{arXiv:2002.08347 [cs, stat]}}
  (\bibinfo{date}{Feb.} \bibinfo{year}{2020}).
\newblock
\urldef\tempurl%
\url{http://arxiv.org/abs/2002.08347}
\showURL{%
\tempurl}
\newblock
\shownote{arXiv: 2002.08347 version: 1.}


\bibitem[\protect\citeauthoryear{Xu, Zhang, Liu, Fan, Sun, Chen, Chen, Wang,
  and Lin}{Xu et~al\mbox{.}}{2019}]%
        {xu_adversarial_2019}
\bibfield{author}{\bibinfo{person}{Kaidi Xu}, \bibinfo{person}{Gaoyuan Zhang},
  \bibinfo{person}{Sijia Liu}, \bibinfo{person}{Quanfu Fan},
  \bibinfo{person}{Mengshu Sun}, \bibinfo{person}{Hongge Chen},
  \bibinfo{person}{Pin-Yu Chen}, \bibinfo{person}{Yanzhi Wang}, {and}
  \bibinfo{person}{Xue Lin}.} \bibinfo{year}{2019}\natexlab{}.
\newblock \showarticletitle{Adversarial {T}-shirt! {Evading} {Person}
  {Detectors} in {A} {Physical} {World}}.
\newblock \bibinfo{journal}{\emph{arXiv:1910.11099 [cs]}} (\bibinfo{date}{Nov.}
  \bibinfo{year}{2019}).
\newblock
\urldef\tempurl%
\url{http://arxiv.org/abs/1910.11099}
\showURL{%
\tempurl}
\newblock
\shownote{arXiv: 1910.11099.}


\bibitem[\protect\citeauthoryear{Zou and Schiebinger}{Zou and
  Schiebinger}{2018}]%
        {zou2018ai}
\bibfield{author}{\bibinfo{person}{James Zou} {and} \bibinfo{person}{Londa
  Schiebinger}.} \bibinfo{year}{2018}\natexlab{}.
\newblock \showarticletitle{AI can be sexist and racist—it’s time to make
  it fair}.
\newblock \bibinfo{journal}{\emph{Nature}}  \bibinfo{volume}{559}
  (\bibinfo{year}{2018}), \bibinfo{pages}{324--326}.
\newblock


\end{thebibliography}
